\documentclass[useAMS,usenatbib]{mnras}
\usepackage{fancyhdr}
\usepackage{graphicx}
\usepackage{color,soul}
\usepackage{amsmath}
\usepackage{amssymb}
\usepackage{times}
\usepackage{newtxmath}
\usepackage{newtxtext}
\usepackage{url}
\newcommand{\HI}{\hbox{\rmfamily H\,{\textsc i}}}
\newcommand{\HIsub}{\hbox{{\scriptsize H}\,{\tiny I}}}
\newcommand{\MHI}{\hbox{$M_{\HIsub}$}}
\newcommand{\rhoHI}{\hbox{$\rho_{\HIsub}$}}
\newcommand{\OHI}{\hbox{$\Omega_{\HIsub}$}}
\newcommand{\msun}{\hbox{${\rm M}_{\odot}$}}
\newcommand{\lsun}{\hbox{${\rm L}_{\odot}$}}
\newcommand{\MgII}{\hbox{\rmfamily Mg\,{\scshape ii}}}
\newcommand{\Ha}{\hbox{\rmfamily H$\alpha$}}
\newcommand{\kms}{\hbox{km\,s$^{-1}$}}
\DeclareRobustCommand{\ion}[2]{\textup{#1\,\textsc{\lowercase{#2}}}}
\newcommand{\0}{\phantom{0}}

\title[${\HI}$ observations of the VVDS~14h field at $z \approx 0.32$]{Neutral hydrogen ({\HI}) gas content of galaxies at $z \approx 0.32$} 

\author[J. Rhee et al.]{Jonghwan Rhee$^{1,2,3}$\thanks{E-mail: jonghwan.rhee@icrar.org}\thanks{Now at ICRAR}, 
  Philip Lah$^{1,4}$, Frank H. Briggs$^{1,3}$, Jayaram N. Chengalur$^{4}$, \newauthor Matthew Colless$^{1,3}$, S. P. Willner$^{5}$, 
  Matthew L. N. Ashby$^{5}$ and Olivier Le F\`{e}vre$^{6}$ \\
$^{1}$Research School of Astronomy and Astrophysics, Australian National University, Canberra, ACT 2611, Australia\\ 
$^{2}$International Centre for Radio Astronomy Research (ICRAR), University of Western Australia, Crawley, WA 6009, Australia\\ 
$^{3}$ARC Centre of Excellence for All-sky Astrophysics (CAASTRO) \\
$^{4}$National Centre for Radio Astrophysics, Tata Institute for Fundamental Research, Pune 411 007, India\\ 
$^{5}$Harvard-Smithsonian Center for Astrophysics, 60 Garden Street, Cambridge, MA 02138, USA \\ 
$^{6}$Aix Marseille Universit\'{e}, CNRS LAM (Laboratoire d'Astrophysique de Marseille), UMR 7326, 13388, Marseille, France}

\begin{document}


\pagerange{\pageref{firstpage}--\pageref{lastpage}} \pubyear{2016}

\maketitle

\label{firstpage}

\begin{abstract}
We use observations made with the Giant Metrewave Radio Telescope (GMRT)
to probe the neutral hydrogen ({\HI}) gas content of field galaxies
in the VIMOS VLT Deep Survey (VVDS) 14h field at $z \approx 0.32$.
Because the {\HI} emission from individual galaxies is too faint to detect at this redshift,
we use an {\HI} spectral stacking technique using the known optical positions
and redshifts of the 165 galaxies in our sample to co-add their {\HI} spectra 
and thus obtain the average {\HI} mass of the galaxies.
Stacked {\HI} measurements of 165 galaxies show 
that $\ga$95~per~cent of the neutral gas is found in blue, star-forming galaxies. 
Among these galaxies, those having lower stellar mass are more gas-rich than more massive ones. 
We apply a volume correction to our {\HI} measurement to evaluate 
the {\HI} gas density at $z \approx 0.32$ as ${\OHI}=(0.50\pm0.18)\times~10^{-3}$
in units of the cosmic critical density. 
This value is in good agreement with previous results at $z < 0.4$, 
suggesting no evolution in the neutral hydrogen gas density over the last $\sim$4~Gyr.
However the $z\approx0.32$ gas 
density is lower than that at $z\sim5$ by at least a factor of two.
\end{abstract}


\section{INTRODUCTION}

A long-standing challenge in galaxy formation and evolution is to address the relationship 
between stars, gas and metals in galaxies. In particular, the mechanisms that supply the fuel 
for star formation in galaxies are unclear. Neutral atomic hydrogen ({\HI}) gas, 
a primary ingredient for star formation, is a key input to understand 
how various processes govern galaxy formation and evolution. 
Cold gas such as neutral atomic hydrogen gas has been only poorly inventoried, 
whereas in contrast astronomers have measured the properties of the stellar components of galaxies 
such as star formation rate density (SFRD) and stellar density over cosmic time \citep[e.g.,][]{Hopkins:2004,Hopkins:2006,Marchesini:2009,Madau:2014} much better.

Two observational approaches have proven to be effective in measuring the abundance of neutral hydrogen in the universe. 
In the high-redshift universe ($z > 2$), saturated Lyman $\alpha$ absorption 
arising from intervening systems seen against bright background sources, viz. the so-called 
damped Lyman $\alpha$ absorption systems (DLAs), measures the abundance of {\HI} gas. 
Recently, large DLA surveys have been used to make high precision measurements of cosmic
{\HI} gas density ({\OHI}) 
\citep[e.g.,][]{Prochaska:2005,Noterdaeme:2009,Noterdaeme:2012,Zafar:2013,Crighton:2015,Neeleman:2016,Sanchez-Ramirez:2016,Bird:2017}. 
These studies show that there exists at least 2 times more hydrogen gas in the high redshift universe than in the local universe. However, this technique is not practical in the low redshift universe 
because the ultra-violet (UV) transition lines of the Lyman series are not observable at $z \la 1.5$ 
with ground-based telescopes and because of the very low incidence rate of DLAs in the local universe. 
This leads to large uncertainties in the low-redshift {\HI} abundance measurements derived from DLAs 
\citep[e.g.,][]{Rao:2006,Meiring:2011}.

In the local universe, the amount of {\HI} gas has been quantified exclusively by {\HI} 21-cm emission. 
In particular, blind {\HI} surveys such as {\HI} Parkes All Sky Survey \citep[HIPASS,][]{Meyer:2004} 
and Arecibo Fast Legacy ALFA \citep[ALFALFA,][]{Giovanelli:2005} have covered a large volume of sky. 
These surveys allow accurate measurements of the local {\HI} mass function and {\HI} mass density of galaxies, 
which are in good agreement \citep{Zwaan:2003,Zwaan:2005,Martin:2010,Haynes:2011}. 
However, this approach has been limited to $z \la 0.2$ 
because the current generation of radio telescopes is too insensitive to detect the feeble {\HI} emission 
from distant populations of galaxies.

Because the two {\HI} measurement techniques work only at the opposite extremes of redshift, 
the {\HI} abundance is poorly constrained at intermediate epochs, $0.2< z <2$. 
This range will be addressed via {\HI} emission studies when the next generation of radio telescopes 
such as ASKAP\footnote{Australian SKA Pathfinder}, FAST\footnote{Five hundred meter Aperture Spherical Telescope}, 
MeerKAT\footnote{The Meer-Karoo Array Telescope} and ultimately the SKA\footnote{Square Kilometre Array} comes online. 
In the interim, some pioneering research programs have used {\HI} spectral stacking techniques to explore this redshift range \citep[e.g.,][]{Lah:2007,Lah:2009,Delhaize:2013,Rhee:2013,Rhee:2016,Kanekar:2016}.

This paper aims to provide an observational constraint on {\HI} gas content in galaxies at $z \approx 0.32$ and its evolution using the Giant Metrewave Radio Telescope (GMRT) and {\HI} spectral stacking techniques. This will help bridge the gap between low and high redshifts, thereby contributing to our understanding of how hydrogen gas has evolved during the last 4~Gyr and the overall cosmic time. 
Section~2 of the paper details the optical data, the optical spectroscopic observations and the data reduction. 
Section~3 describes the radio data. The main results of our {\HI} spectral stacking experiment are presented in Section~4, 
and the relationship between galaxy stellar mass and {\HI} mass is investigated in Section~5. 
Finally, the cosmic evolution of {\HI} gas density is presented in Section~6, followed by a summary in Section~7. 
We adopt the concordance cosmological parameters of $\Omega_{\Lambda}=0.7$, $\Omega_{M}=0.3$ and 
$H_{0}=$~70~\kms~Mpc$^{-1}$ throughout this paper.

\begin{figure*}
 \centering
 \includegraphics[width=160mm]{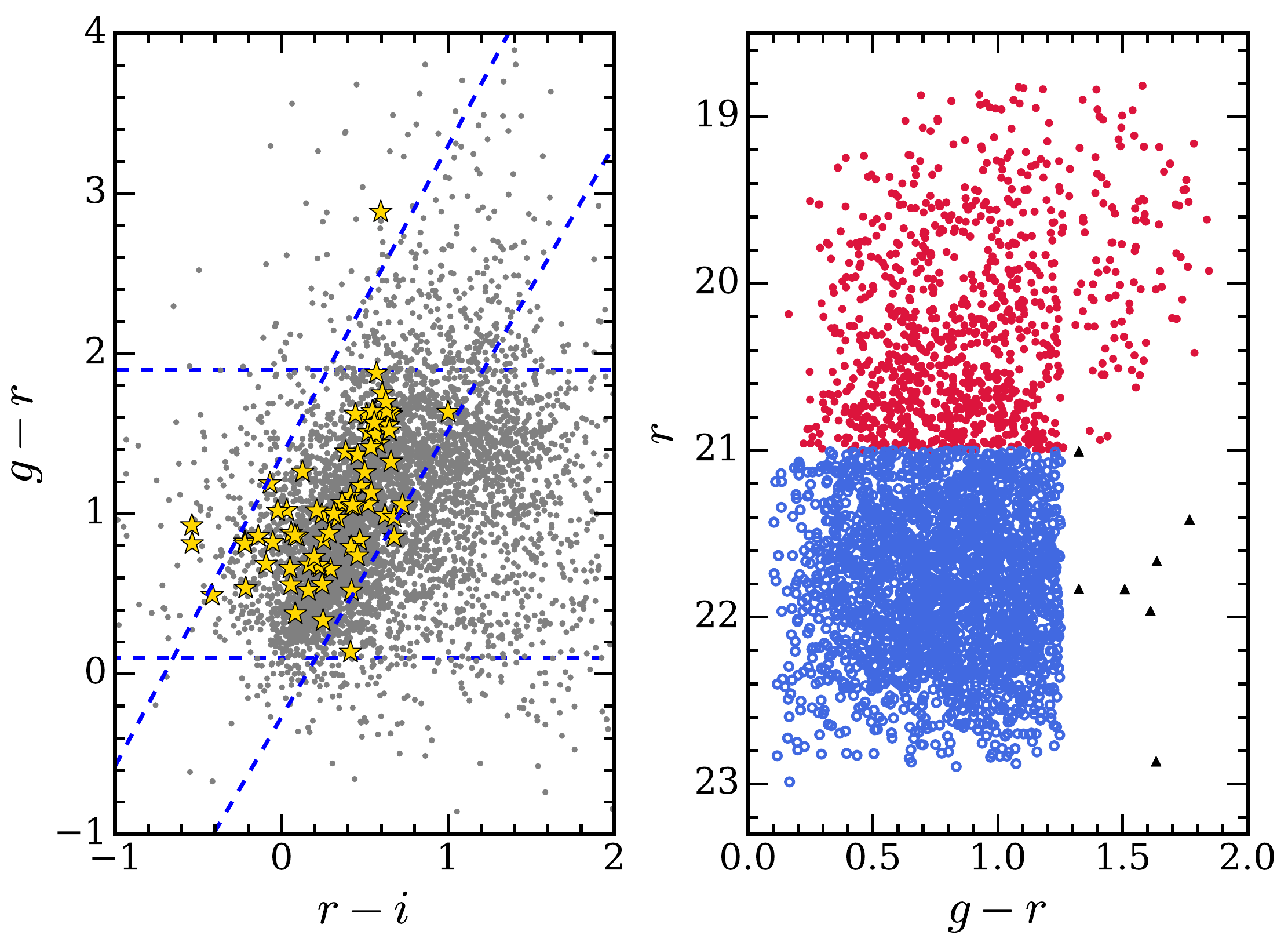}
 \caption{ {\it left:} Colour--colour diagram for the VVDS~14h field.  
   Yellow stars show objects in the VVDS catalogue having $0.30 < z < 0.34$, 
   and gray points show catalogue objects outside this redshift range. 
   Dashed lines show the colour selection criteria adopted to select targets for the MMT and AAT observations. 
   {\it right:} Colour--magnitude diagram for galaxies selected for spectroscopy. 
   The bright sample is shown by red closed circles and the  faint blue sample by blue open circles. 
   The gap in the bright sample at $g-r \approx 1.25$ reflects galaxy colours, not sample selection. 
   Faint, red objects (black triangles) had pre-existing VVDS spectra.}
 \label{fig:colour_cut} 
\end{figure*}

\section{Optical Data}

\subsection{Target Field Selection} 

We selected one of three independent fields in the VIMOS VLT Deep Survey (VVDS)-Wide \citep{Le-Fevre:2005, Garilli:2008,Le-Fevre:2013}, 
VVDS~14h, as the target field for {\HI} 21-cm observations with the GMRT\null. 
The VVDS-Wide was designed to measure redshifts of galaxies with a limiting magnitude of ${\it I_{AB}} = 22.5$ 
out to $z \sim 1$ using the VIsible Multi-Object Spectrograph (VIMOS) 
on the European Southern Observatory (ESO) Very Large Telescope (VLT). 

For our observations, the GMRT frequencies restricted the redshift range to $0.30 < z < 0.34$.
In this limited redshift interval and field-of-view, there were only 41 reliable redshifts 
in the VVDS-Wide catalogue available for the radio data analysis. 
This is an insufficient number of redshifts to obtain a strong {\HI} 21-cm signal using the stacking technique.
Moreover, the VVDS has a low spectral resolution of $R=230$ with a dispersion 7~{\AA}~pixel$^{-1}$, 
resulting in a large redshift uncertainty of $\sim$300~\kms\ \citep{Le-Fevre:2005}. 
This increases the uncertainty in the stacked {\HI} signal and the subsequent measurements. 
To obtain more redshifts with higher precision, we used two optical telescopes with multi-object optical spectrographs: 
Hectospec \citep{Fabricant:1998,Fabricant:2005} on the 6.5-m MMT and 
AAOmega \citep{Saunders:2004,Smith:2004,Sharp:2006} on the 3.9-m Anglo Australian Telescope (AAT).

\subsection{Sample Selection}

The effective field of view for our optical spectroscopic observations corresponds to a circular field of one degree diameter,
smaller than the total area of 2.2~deg$^{2}$ surveyed for the VVDS~14h field in the original VVDS-Wide. 
This field has good optical imaging coverage in both $R$ and $I$ bands along with Sloan Digital Sky Survey (SDSS) imaging, 
and the SDSS archive contains 16,881 possible spectroscopic targets. 
This number is too large to reasonably observe with Hectospec and AAOmega, 
and many of the targets are too faint to obtain redshifts with these instruments.
However, using the existing photometry, one can apply the SDSS star--galaxy criteria 
and a series of colour cuts to select objects likely to be galaxies in our desired redshift range.

Comparing SDSS magnitudes and redshifts obtained from the VVDS project, 
we investigated various colour distributions of galaxies in the VVDS~14h field. 
Fig.~\ref{fig:colour_cut} (left) shows the colour--colour locus of objects in the redshift range covered by the GMRT observations. 
The $g-r$ vs. $r-i$ colour cuts indicated in the figure were used to make target lists of $r < 21$ galaxies 
for the MMT and AAT spectroscopic observations. 
For galaxies with $r > 21$, an additional requirement of $g-r \leq 1.25$ was imposed 
because spectroscopy of such faint galaxies is unlikely to yield redshifts
unless the galaxy has emission lines, and blue galaxies are more likely to have emission lines.

The bright galaxies were assigned the highest priority for observing because they were more likely to yield redshifts. 
Applying the selection criteria, 2240 target galaxies were selected for Hectospec observations, 
and a target list of 2161 galaxies was made for AAOmega observations. 
Included in the lists were 41 galaxies that already had VVDS redshifts but had large uncertainties. 
Seven of these are red and faint and would not otherwise have been included in the sample. (See right panel of Fig~\ref{fig:colour_cut}.)

\begin{figure*}
 \centering
 \includegraphics[width=150mm]{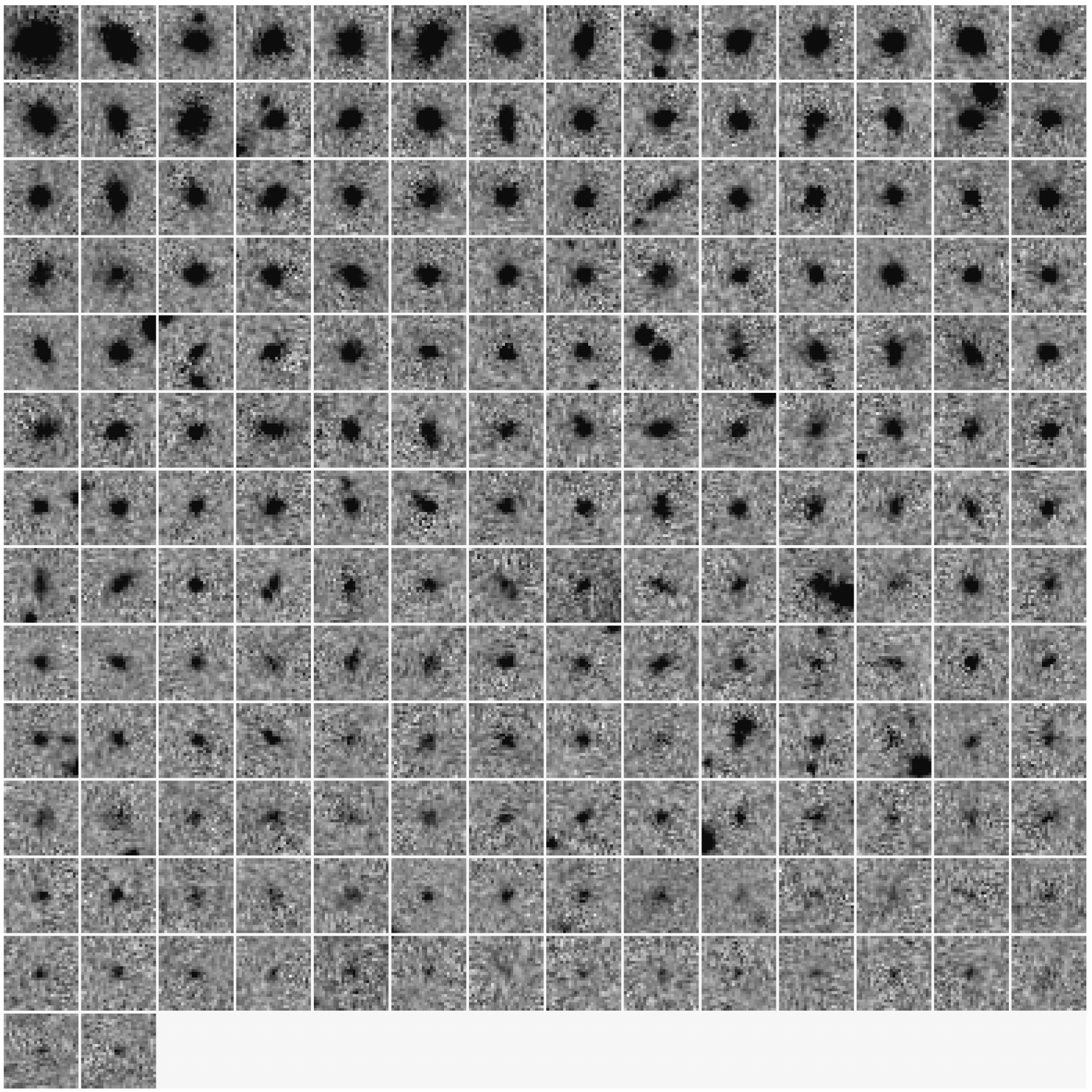}
 \caption{SDSS $r$-band thumbnail images of 184 sample galaxies in the VVDS~14h field 
   that have $0.30 < z < 0.34$ from Hectospec and AAOmega spectroscopic observations 
   and lie within the GMRT beam coverage. The size of each thumbnail image is 10{\arcsec} by 10{\arcsec}. 
   They are ordered by increasing $r$ magnitude.} 
 \label{fig:tile_sample}
\end{figure*}

\subsection{Observations and Data Reduction} \label{sec:obs}

Hectospec and AAOmega are ideal spectrographs for efficiently obtaining a large number of galaxy redshifts 
because they have large fields of view and many fibres to simultaneously obtain spectra. 
The Hectospec instrument has 300 fibers and a field of view one degree in diameter. 
Wavelength coverage is 3700--9200~{\AA} at 1.2~{\AA}~pixel$^{-1}$ and 
$R \sim 1000$--$2000$ \citep{Fabricant:1998,Fabricant:2005}.
MMT observations of six fibre configurations were carried out in queue observing mode over 5 nights 
between February and April 2011 with a 1.5~hr exposure time 
(divided into three 0.5~hr exposures for cosmic ray removal) for each fibre configuration. 
1526 of 2240 target galaxies, corresponding to $\sim$68\% of our sample, were successfully observed.

The Hectospec spectra were processed through a semi-automatic pipeline developed specifically for Hectospec 
and based on {\sc iraf} scripts \citep{Mink:2007}. 
It includes standard calibration with bias, dark and flat-fielding followed by wavelength calibration with comparison lamp exposures, 
then followed by subtracting night-sky emission lines. 
After sky lines were removed, each spectrum was cross-correlated against a set of templates of various galaxy spectra 
using {\sc rvsao} \citep{Kurtz:1992, Kurtz:1998}, an {\sc iraf} add-on package developed at the Smithsonian Astrophysical Observatory 
(SAO)\footnote{http://tdc-www.harvard.edu/iraf/rvsao/}. 
This procedure yielded redshifts with quality 3 or better for 1296 galaxies.  
The quality flags are defined by visual inspection as numbers from 5 to 0:
\begin{description}
  \item[\it flag 5:] 100\% convincing redshift with high SNR multiple emission or absorption lines
  \item[\it flag 4:] Reliable redshift with a mixture of strong and weak spectral lines 
  \item[\it flag 3:] Probable redshift showing a couple of spectral lines with low SNR
  \item[\it flag 2--0:] Rejected.
\end{description}

The spectra of target galaxies unobserved by the MMT were taken using the AAOmega, 
a dual-beam and multi-fibre spectrograph mounted on the 3.9-m AAT at Siding Spring Observatory (SSO). 
AAOmega allows for the simultaneous observation of 392 targets including science objects, 
sky positions and guide stars over a two-degree field using the 2dF fibre positioner \citep{Lewis:2002}. 
AAOmega provides $R=1300$ with a dispersion of 1{\AA}~pixel$^{-1}$ 
\citep{Saunders:2004,Smith:2004,Sharp:2006} using the 5700~\AA\ dichroic.

The AAT observations were done on 2011 June 1--4 in the wavelength range 3700~{\AA} to 9000~{\AA} 
with an exposure time of 2.5~hr for each of four fibre configurations. 
The data were reduced using dedicated software developed at the Australian Astronomical Observatory 
(AAO)\footnote{http://www.aao.gov.au/2df/aaomega/aaomega\_2dfdr.html} called {\sc 2dfdr} \citep{Croom:2004,Sharp:2010}. 
{\sc 2dfdr} applies a standard sequence of data reduction to extract a 1D spectrum of each object from 2D images taken from 
AAOmega: bias subtraction, flat-fielding, fibre trace fitting and wavelength calibration. 
Then the redshifts were derived with the interactive cross-correlation software package {\sc runz} \citep{Saunders:2004b}, 
assigning the same redshift quality flags as used for Hectospec redshifts. 
The spectra yielded 344 reliable redshifts out of 1398 galaxies targeted. 
The success rate was lower than that of the MMT/Hectospec because bright targets had already been observed by Hectospec, 
and the AAOmega targets were therefore fainter, and the weather at the telescope was poor.

\subsection{Redshift Measurements}

\begin{figure}
 \centering
 \includegraphics[width=\columnwidth]{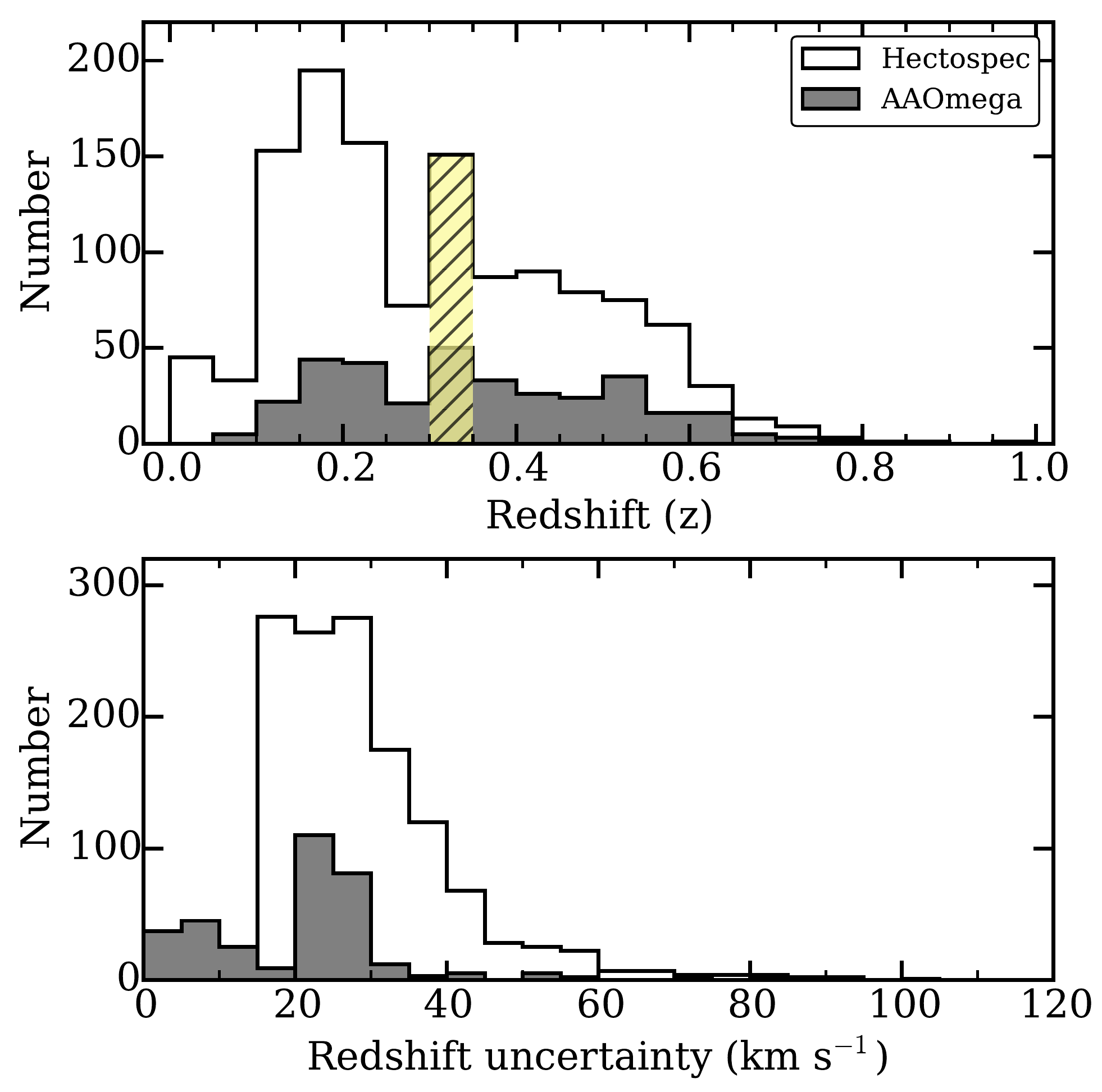}
 \caption{Redshift ({\it top}) and redshift uncertainty ({\it bottom}) distributions of the 1640 galaxies in the VVDS~14h
   field with redshifts obtained from MMT/Hectospec and AAT/AAOmega. 
   The GMRT observing frequency corresponds to the yellow-coloured, hatched interval of redshift.} 
 \label{fig:redshift_all}
\end{figure}  

\begin{table*}
 \caption{Redshift catalogue of the VVDS~14h field taken from the AAT and MMT observations. 
   The full table in machine-readable form is available online.} 
 \centering
 \label{tab:zcat}
 \begin{tabular}{@{}cccccc}
   \hline
     SDSS Object ID &  R.A. (J2000)  &  Dec. (J2000)  &  Redshift ($z$) & Quality flag & Instrument \\
     \hline
     588010880382534322  &  13:57:36.12  &  5:31:47.7  &  0.320044~$\pm$~0.000065  &  4  &  Hectospec \\
     587729159509639373  &  13:57:05.62  &  4:49:05.6  &  0.320850~$\pm$~0.000095  &  4  &  Hectospec \\
     588010880382534003  &  13:57:29.44  &  5:29:08.1  &  0.321540~$\pm$~0.000080  &  4  &  AAOmega  \\
     587729160046707183  &  13:59:05.01  &  5:14:19.2  &  0.322960~$\pm$~0.000090  &  4  &  AAOmega  \\
     588010879845793833  &  13:58:46.00  &  5:06:15.9  &  0.322980~$\pm$~0.000080  &  4  &  AAOmega  \\
     588010879308792391  &  13:57:41.24  &  4:41:23.2  &  0.323292~$\pm$~0.000139  &  4  &  Hectospec \\
     588010879308792194  &  13:57:33.63  &  4:43:08.7  &  0.323530~$\pm$~0.000080  &  4  &  AAOmega  \\
     588010879308792394  &  13:57:41.28  &  4:41:21.2  &  0.323720~$\pm$~0.000184  &  3  &  Hectospec \\
     587729159509639598  &  13:57:04.37  &  4:52:56.9  &  0.324770~$\pm$~0.000080  &  4  &  AAOmega  \\
     588010879308923480  &  13:59:05.97  &  4:44:26.6  &  0.325410~$\pm$~0.000112  &  4  &  Hectospec \\
     587729159509901597  &  13:59:15.32  &  4:45:24.9  &  0.325543~$\pm$~0.000081  &  4  &  Hectospec \\
     587729159509901676  &  13:59:15.91  &  4:45:58.9  &  0.325547~$\pm$~0.000068  &  5  &  Hectospec \\
  \hline
 \end{tabular}
\end{table*}

Redshift measurements for both instruments are based on cross-correlation methodology \citep{Tonry:1979}. 
The calculated redshift uncertainties include statistical and systematic effects: 
noise, calibration errors, and template mismatching. \citet{Tonry:1979} and \citet{Kurtz:1998} give more detail.
As seen in Fig.~\ref{fig:redshift_all}, redshift measurements from both observatories have mean uncertainty of $\sim$30~\kms. 
This is important for {\HI} stacking because it narrows the velocity window needed for co-adding the {\HI} spectra.

The combined observations give a total of 1640 galaxy redshifts.\footnote{Nine galaxies were observed at both telescopes, 
and their redshifts agree.} (See Table~\ref{tab:zcat}.) 
Of these, 184  or 11\% are within the GMRT wavelength range corresponding to $0.30 < z < 0.34$. Fig.~\ref{fig:tile_sample} shows the appearance of these galaxies.

Subsequent to our observations, the VVDS project released the final catalogues of its surveys \citep{Le-Fevre:2013}. 
The VVDS~14h field is included in the VVDS-Wide survey, 
and the catalogue has 60 objects within the GMRT beam and frequency coverage. 
Of these, 15 galaxies have reliable (based on its own quality flag, 
${\it zflag} \ge 3$) redshifts not already known from our
observations.\footnote{An additional 11 galaxies have 
VVDS redshifts with low quality and are not included.} 
The addition of redshifts from the final VVDS catalogue increased the total number of galaxies available to be co-added to 199.

\subsection{Galaxy Classification} \label{sec:galaxy_class}

\begin{figure}
 \centering
 \includegraphics[width=\columnwidth]{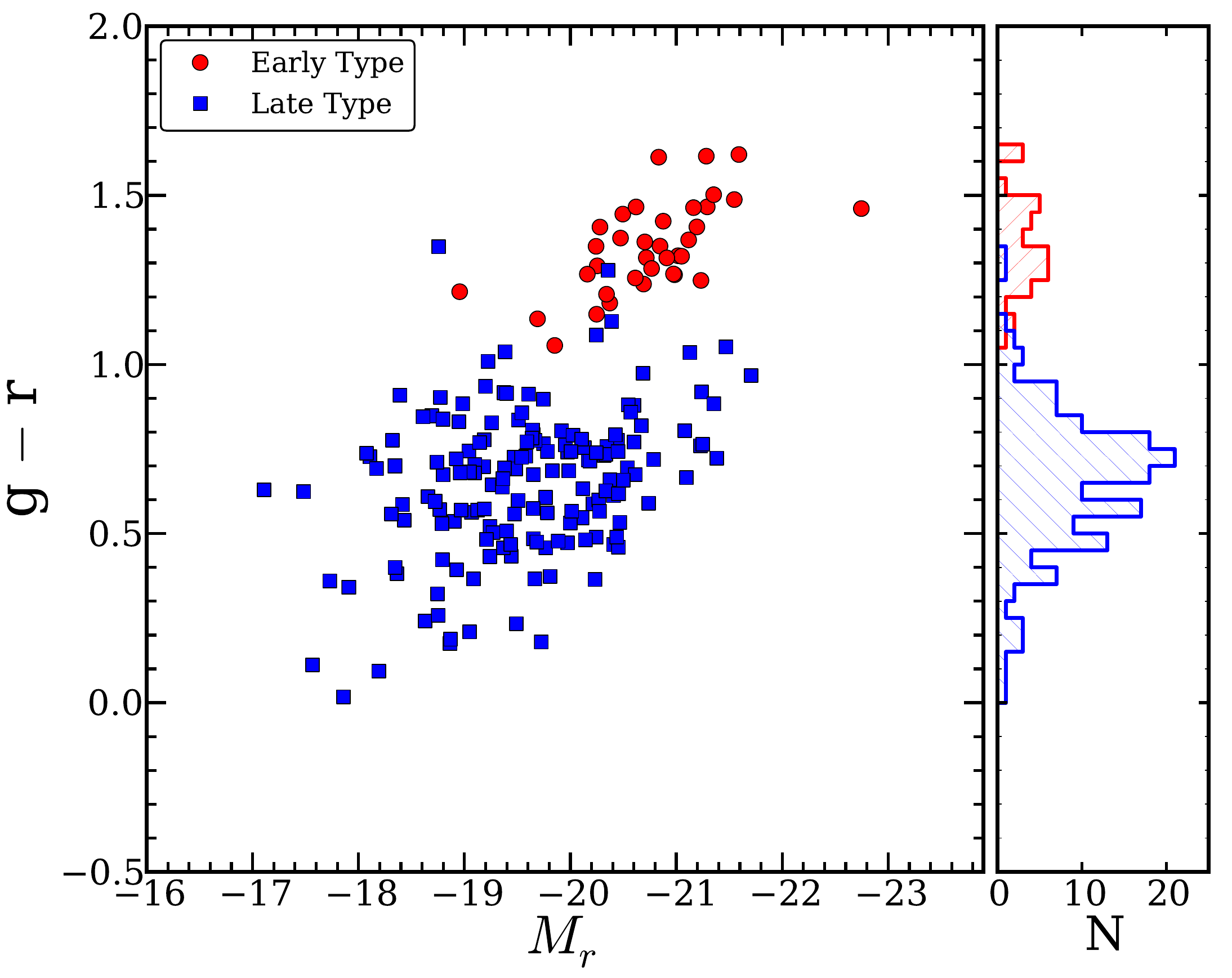}
 \caption{{\it left:} Colour--magnitude diagram of the 199 galaxies in the VVDS~14h field with $0.30 < z <0.34$. 
   Red circles denote objects spectroscopically classified as early-type and blue squares those classified as  late-type. 
   {\it right:} Histograms showing the $g-r$ colour distribution. 
   Red hatching ($/$) denotes early types and blue hatching ($\backslash$) late types.}
 \label{fig:cmd_vvds}
\end{figure}  

Galaxies with optical redshifts in the GMRT range were classified into two subsamples---early type 
and late type---based on the dominance of absorption or emission lines.  
Initial classification was based on the  templates used during the cross-correlation procedure for deriving redshifts, 
and all classifications were verified by visual inspection of the spectra. 
Galaxies with emission lines such as [\ion{O}{ii}] and {\Ha}, i.e., typical emission lines seen 
in spectra of star-forming galaxies, were classified as late-type. 
Those with strong absorption lines, e.g., Ca H and K, without emission lines were classified as early-type.
Of the 184 galaxies observed by the AAT and the MMT, 34 were classified as early-type and 150 galaxies were
classified as late-type. Visual inspection of the additional spectra from the VVDS-Wide catalogue added  2 early-type 
and 13 late-type galaxies for a total of 36 early-type and 163 late-type galaxies available for {\HI} stacking.

All sample galaxies have photometric data from the SDSS, including $ugriz$ magnitudes. 
Based on the SDSS photometry of our sample, galaxies that were classified by spectrum 
were cross-checked with their colours \citep{Baldry:2004}. 
For this, the SDSS photometry of all galaxies was corrected for the effect of dust extinction from the Milky Way 
using the dust maps of \citet{Schlegel:1998}.\footnote{The maximum correction was 0.17~magnitude.} 
The k-correction \citep{Blanton:2003,Blanton:2007} was then applied to convert all observed magnitudes 
to rest-frame magnitudes.\footnote{Mean corrections in individual bands were 0.35--0.77~magnitudes.} 
As shown in Fig.~\ref{fig:cmd_vvds}, all galaxies classified as early type have $g-r >1.0$, 
while most late-type galaxies have $g-r <1.0$. 
This is the usual colour bimodality: early-type galaxies are located in the `red sequence' region, 
and most late-type galaxies are distributed in the area known as the `blue cloud' with a few showing redder colours because of dust. 
For the analysis in this paper, we used the spectroscopic classifications.

\section{Radio Data}

\subsection{Observations}

\begin{figure*}
 \centering
 \includegraphics[width=150mm]{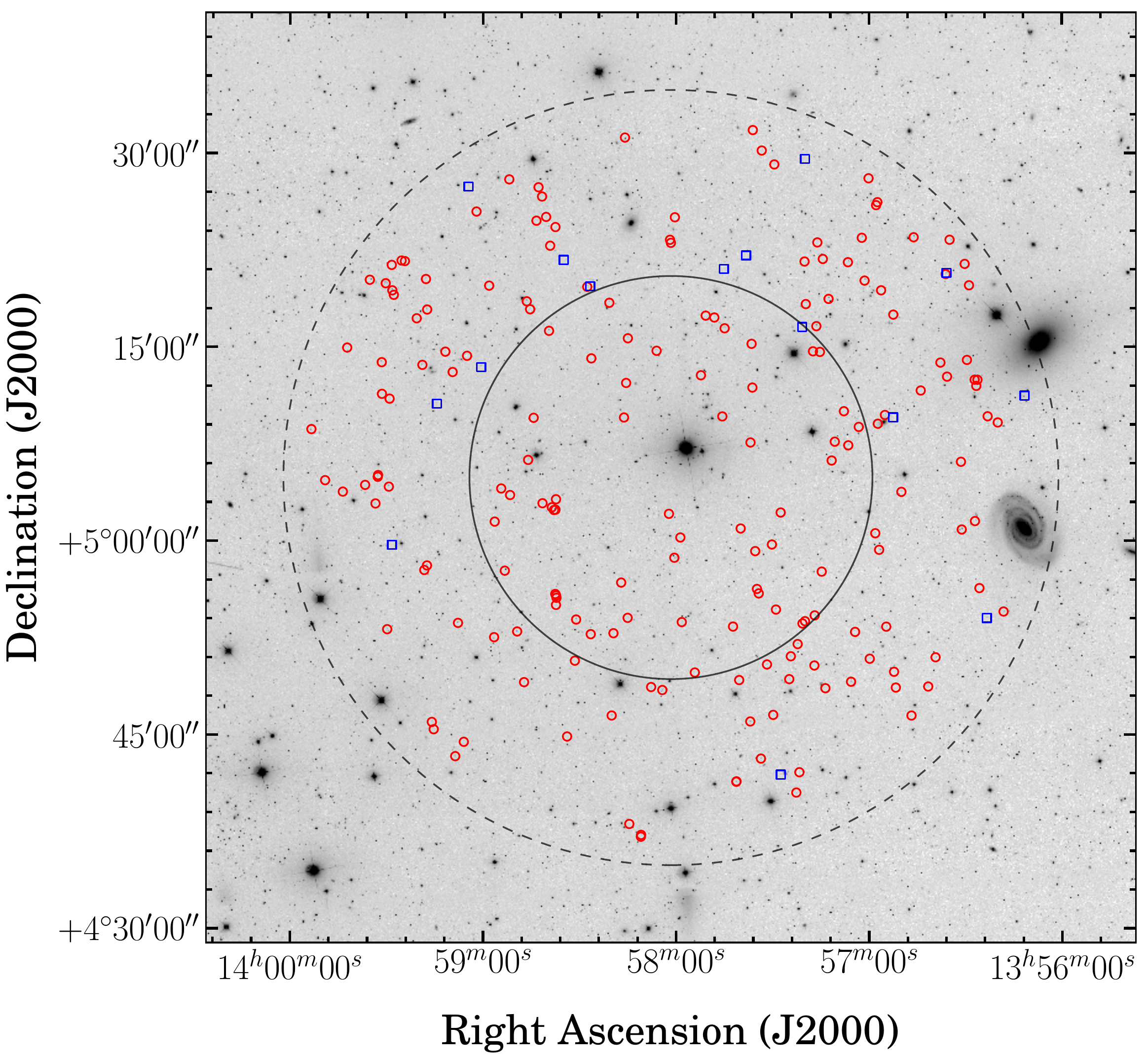}
 \caption{SDSS $r$-band negative image of the VVDS~14h field.  
   Red circles mark galaxies with  $0.30 < z < 0.34$ measured from our observations. Blue squares mark galaxies within the same redshift range measured by the VVDS-Wide survey. 
   The central solid circle corresponds to the primary beam diameter (HPBW) of 31.2~arcmin at 1076~MHz. 
   The outer dashed circle, 57\arcmin\ in diameter, represents the 10\% level of the GMRT beam. }
 \label{fig:gmrt}
\end{figure*}

The GMRT consists of 30 steerable parabolic dishes each 45~m in diameter \citep{Swarup:1991}. 
Over the central ${\sim}1$~km, 14 antennas are deployed in a random pattern, 
and the remaining 16 antennas are spread out along three outer arms, forming a `Y'-shaped configuration. 
The shortest baseline in the central square is ${\sim}100$~m, and the longest separation between antennas is about 25~km. 
This hybrid antenna configuration has good $uv$ coverage and sensitivity, 
providing the GMRT with good imaging performance and high angular resolution.  

The GMRT {\HI} observations of the VVDS~14h field were carried out over a total of 21 days from 2008 to 2011. 
The total integration time is $\sim$136~hr. Excluding flux and phase calibrator scans, 117~hr were used for on-source integration.  
3C~147 and 3C~286 were observed for primary flux calibration, 
and a VLA calibrator source PKS 1345+12 was used as a phase calibrator. 
The observations of PKS 1345+12 gave a measured flux density of 6.01$\pm$0.01~Jy.

The old hardware correlator of the GMRT had a total bandwidth of 32~MHz, split into two 16~MHz sidebands. 
A new software correlator replaced the old hardware correlator in 2010 \citep{Roy:2010}. 
The new software correlator has one continuous observing band with 33~MHz bandwidth. 
We took data with the old hardware correlator in 2008 and 2009. 
Data were obtained with the new software correlator in 2011. 
Data from the old and the new correlator systems amount to 74~per~cent ($\sim$100~hr) and 26 per~cent ($\sim$36~hr), respectively. 
The two 16~MHz-wide sidebands of the old hardware correlator observations cover the frequency range from 1092~MHz 
to 1060~MHz, corresponding to a redshift range of $0.30 < z < 0.34$. 
Each sideband has two polarisations and 128 spectral channels, giving a channel width of 0.125~MHz. 
Data taken from the new correlator system in 2011 have a single 33~MHz-wide bandwidth with two polarisations 
and 256 spectral channels of 0.130~MHz channel spacing covering the same frequency range as the old correlator. 
  
\begin{figure}
 \centering
 \includegraphics[width=\columnwidth]{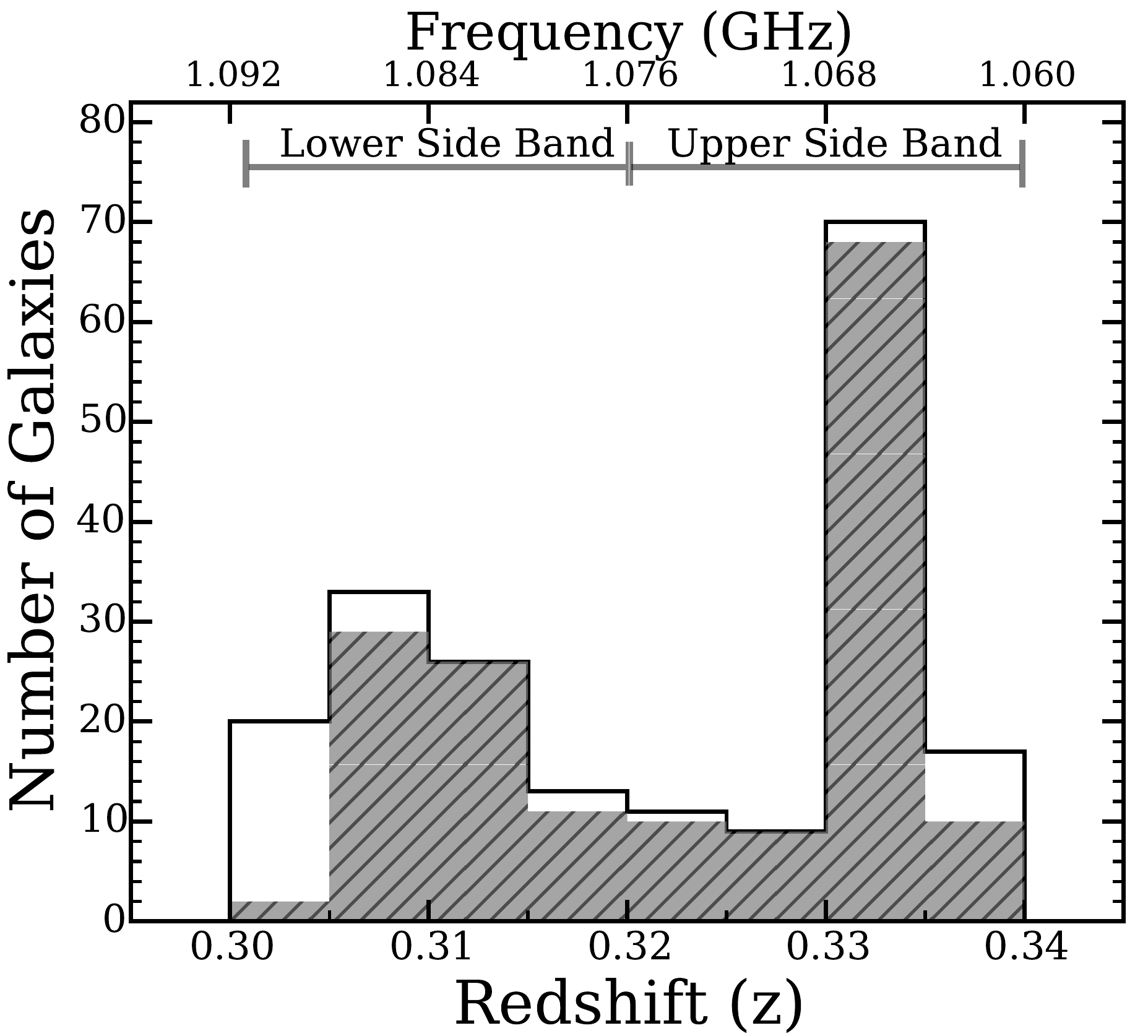}
 \caption{Redshift distribution of the VVDS~14h field galaxies that lie within the GMRT frequency range. 
   The grey histogram with hatching shows the redshifts used for the final {\HI} stacking analysis after dropping targets affected by interference. The upper horizontal axis represents the frequency corresponding to the redshifts.} 
 \label{fig:redshift_hist}
\end{figure}

The pointing centre of the VVDS~14h field observations was R.A.~13$^{\rm h}$58$^{\rm m}$01$\rm \fs$60, 
Dec.~+05\degr04\arcmin54\farcs0 (J2000). 
Fig.~\ref{fig:gmrt} shows the GMRT beam coverage of the VVDS~14h field along with the positions of 
the catalogued galaxies with redshifts in the GMRT frequency band. 
Galaxies within the 10~per~cent beam level were used for the subsequent {\HI} spectral stacking analysis.
The redshift distribution of the stackable galaxies within the GMRT frequency range is shown in Fig.~\ref{fig:redshift_hist}. 
46~per~cent of the total sample is included in the lower sideband while the rest lies in the upper sideband of the old correlator.  

\subsection{Data Reduction}

\subsubsection{Calibration}
Data reduction was conducted primarily with the Common Astronomy Software Applications 
\citep[{\sc casa},][]{McMullin:2007}\footnote{http://casa.nrao.edu} following standard procedures: 
flagging, calibration, self-calibration and imaging. 
Before implementing {\sc casa} for calibration and subsequent processes, 
we ran an automated flagging and calibration software developed for the GMRT data, 
the so-called {\sc flagcal}, to automatically search and discard corrupted data
caused by malfunctioning antennas and radio frequency interference (RFI) \citep{Prasad:2012,Chengalur:2013}. 
Assuming that true visibilities vary smoothly with time and frequency, the corrupted data 
due to RFI and instrumental errors can be identified by their deviant behaviours from the expected smoothness. 
To apply this approach, the data should be corrected for different antenna gains. 
{\sc flagcal} carries out calibration and flagging in an iterative way: calibration parameters were derived and applied to
the raw data, and then the data were inspected to find and remove corrupted data, 
followed by iterations of this procedure of calibrating and flagging with increasingly stringent criteria.

\begin{figure}
 \centering
 \includegraphics[width=\columnwidth]{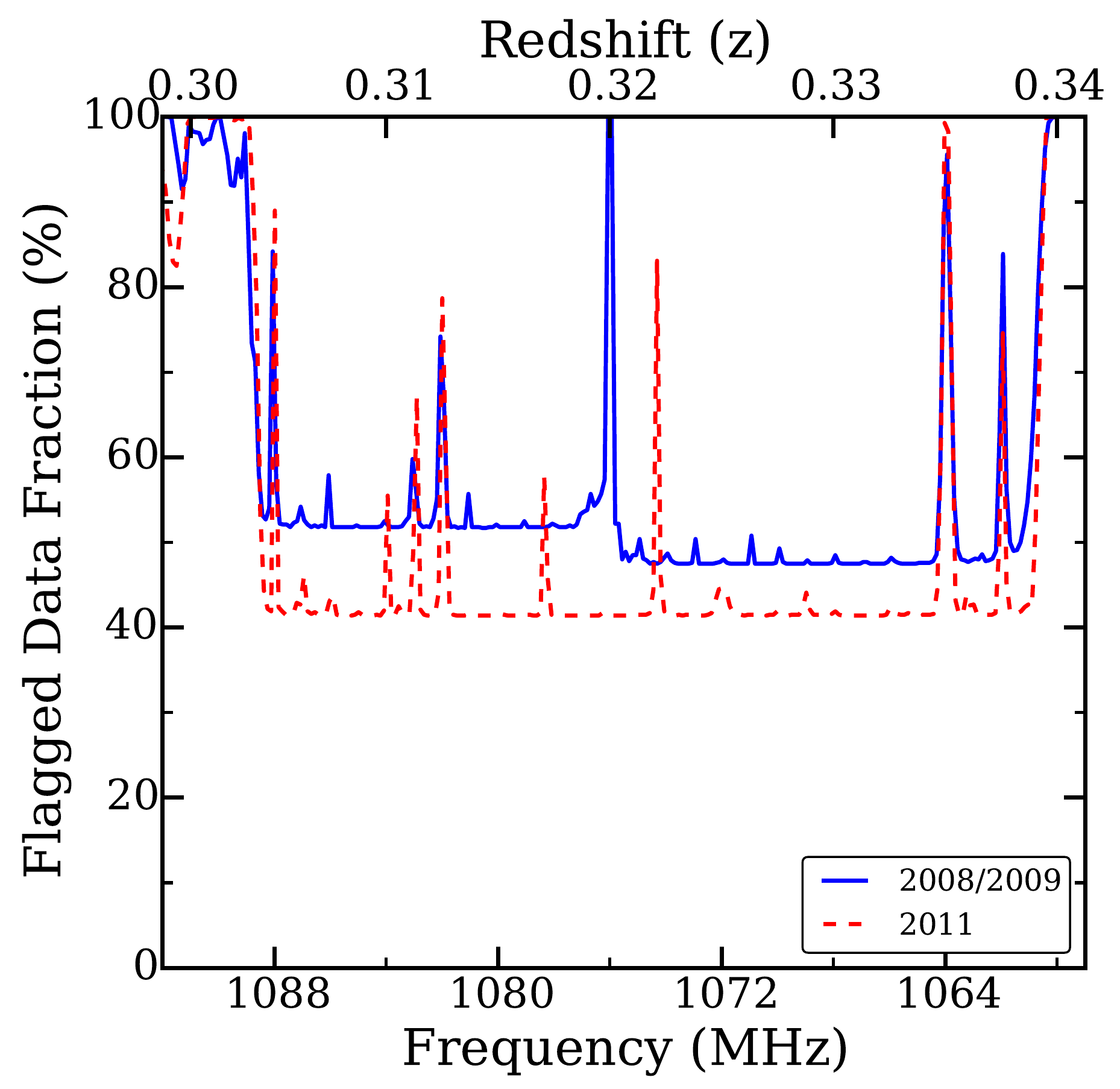}
 \caption{Flagged data fraction of the GMRT data as a function of frequency channel. 
   The blue solid and the red broken lines denote data taken in 2008--2009 and in 2011, respectively.} 
 \label{fig:rfi}
\end{figure}

After editing away bad data via {\sc flagcal}, the visibilities were imported into {\sc casa} for the main data reduction procedures. 
Owing to the different frequency configurations, the pre-2011 and 2011 data sets were processed separately 
although the same data reduction procedure was used. 
Each of the 21 days was individually calibrated, keeping separate the two sidebands in the 2008 and 2009 data. 
Then all datasets were concatenated for computation of the final continuum images and data cubes. 
The data were inspected again using {\sc casa} visualisation tools, and additional flagging was performed as required. 
For calibration, bandpass calibration was first determined with flux and phase calibrators, 
and then calibration for flux and phase with 3C~147, 3C~286 and PKS~1345+12 was done with the central 80 channels of a total
128 channels in each sideband, avoiding channels severely affected by RFI\null. 
Science data to which all calibration solutions had been applied were split and inspected again 
followed by continuum imaging to check whether there remained bad data that could cause artefacts in the continuum image. 
After additional bad data found from this inspection were flagged, each data set was self-calibrated 
with six bright radio continuum sources in the VVDS~14h field to diminish residual amplitude and phase errors.

To make continuum images of each sideband for self-calibration, we used 100 out of 128 channels in each sideband. 
These were averaged in groups of 10 adjacent channels to form 10 sub-bands of 1.25 MHz each to be used in 
gridding in the $uv$ plane in preparation for imaging. 
These bands are sufficiently narrow to avoid bandwidth smearing across our field of view. 
For self-calibration, a total of 4 loops were repeated in each sideband: 
2 phase self-calibration loops and 2 amplitude $\&$ phase self-calibration loops. 
All self-calibrated 2008--2009 data were concatenated based on sideband, and self-calibration was carried out again 
with the entire combined data to ensure consistency over all the data observed on different days. 
For 2011 data, the data reduction sequence was the same as the 2008--2009 data: initial editing with {\sc flagcal}, 
calibration, self-calibration, concatenation and self-calibration.

The data used in the final reduction were inspected to check how much data have been flagged due to RFI. 
Fig.~\ref{fig:rfi} shows the frequency range observed by the GMRT is severely affected by RFI and instrumental failure. 
While more than 50~per~cent of the data taken in 2008 and 2009 were flagged, 
the 2011 data were somewhat less contaminated by RFI, losing $\sim$42~per~cent to flagging. 
The upgrade of the GMRT backend systems to the software correlator likely contributes to this improvement in RFI mitigation. 
Having 50~per~cent of the data remaining after all flagging is equivalent to having 21 of the 30 GMRT antennas 
working perfectly for the whole observation.

\begin{figure}
 \centering
 \includegraphics[width=\columnwidth]{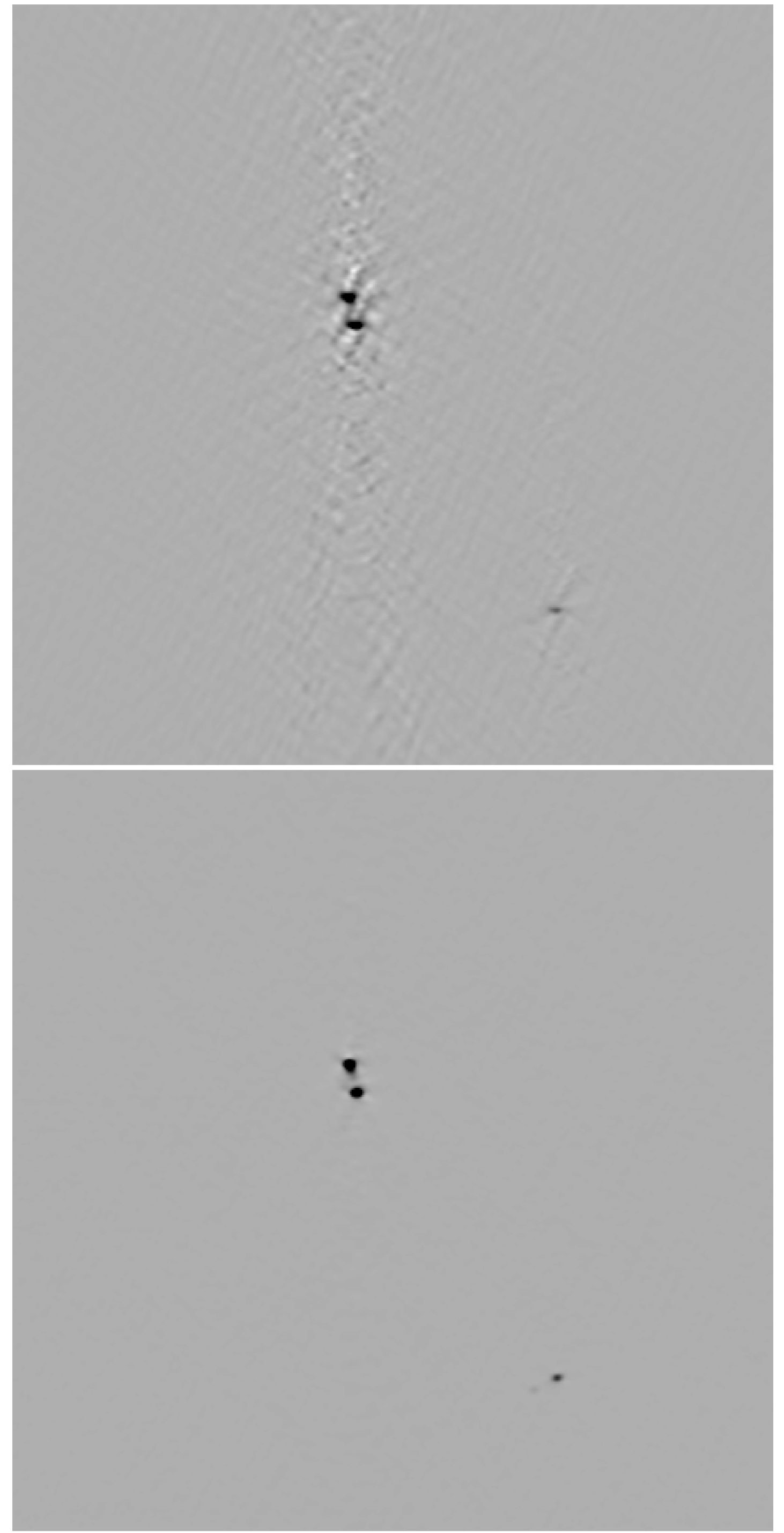}
 \caption{Negative continuum images of bright sources in the VVDS~14h field 
   before ({\it top}) and after ({\it bottom}) applying $w$-projection. 
   The upper sources and lower sources in each panel are $\sim$13.3 and $\sim$16~arcmin away 
   from the observing centre of the GMRT, respectively. 
   The image size of each panel is 6~arcmin on a side. The same intensity scale was applied to the two images.}
 \label{fig:wproj}
\end{figure}

\subsubsection{Continuum map}
The final continuum map has a $\sim$1~deg field-of-view (FoV) 
extending out to the 10 per~cent beam level of the GMRT primary beam. 
To make the map we concatenated all data including the 2011 observations. 
In order to avoid bandwidth smearing, double sideband data used 100 channels with averaging every 10 channels, 
and the single band data were also averaged with every 10 of 190 channels. 
The pixel size for imaging was 0\farcs85, and the `{\sc robust}' weighting parameter \citep{D_Briggs:1995} was set to zero, 
corresponding to a compromise between natural and uniform weighting.

When making continuum images as well as spectral data cubes at arcsecond resolution, 
the one degree FoV of the VVDS~14h field is large enough to require a wide-field imaging algorithm. 
Unlike one-dimensional interferometric arrays such as the WSRT, the GMRT is two-dimensional and has non-coplanar baselines. 
The effect of the non-coplanar line-of-sight components ($w$-term) of the interferometer baselines has been one of 
the factors limiting wide-field imaging quality though it can be ignored for small-field imaging. 
However, the effect is no longer negligible as the imaging size required by current and future radio telescopes increases. 
To deal with the effect of the $w$-term on wide-field imaging, there are currently two commonly used algorithms: 
faceting \citep{Cornwell:1992} and $w$-projection \citep{Cornwell:2008}. 
The {\sc casa} imager is capable of implementing both algorithms.

For continuum imaging and creating the spectral data cube of the VVDS 14h field, 
the $w$-projection algorithm was applied in {\sc casa} 
because this has superior performance in speed and dynamic range to the faceting algorithm \citep{Cornwell:2008}. 
Fig.~\ref{fig:wproj} shows continuum images of bright sources $\ga$13 arcmin away from the phase centre. 
As seen in the top panel, the $w$-term effect results in errors around bright sources in the continuum image. 
Applying the $w$-projection algorithm, these disappear as seen in the bottom panel, 
and the peak signal-to-noise ratio (SNR) of the brightest source in the image increases by 4.7 times. 
The weaker, distorted sources to the lower right in the images are recovered cleanly in the bottom panel by applying $w$-projection.

Using bright sources with high SNR in the continuum maps, the astrometric accuracy of the GMRT data was checked 
by matching them against their positions in the VLA FIRST survey catalogue \citep{Becker:1995}. 
The offset between the two coordinates is $\sim$0.48~arcsec on average which is less than the image pixel size of 0.85~arcsec. 
The maximum measured offset for a single object was 1~arcsec, 
which is smaller than the GMRT synthesised beam size of $3.4\arcsec \times2.3\arcsec$. 
The root-mean-square ({\it rms}) noise level of the continuum map is 15.07~$\mu$Jy~beam$^{-1}$.

\subsubsection{Spectral data cube} 
To obtain the final spectral data cubes, the radio continuum has to be subtracted from the spectral $uv$ data. 
Two methods were used sequentially to achieve this end. 
Firstly, clean component models of continuum sources were directly Fourier-transformed and 
then subtracted from the $uv$ domain using the {\sc casa} task \texttt{`uvsub'}. 
After creation of the spectral image cube by transforming the visibilities for each channel to its image plane,
the small residual continuum flux that remained after the clean components subtraction was removed 
by subtracting a linear fit to the spectrum for each spatial pixel of the cube 
using the {\sc casa} task \texttt{`imcontsub'}. That is, a spectral data cube was made from the continuum-subtracted $uv$ data, 
and then the linear fit was applied to the data cube to measure and subtract the residual continuum flux. 
Three final data cubes were made: two-sideband (LSB and USB) data cubes each with 128 channels from 2008--2009 data 
and one data cube with 256 channels from the 2011 single-band (GSB) data. 
Each data cube has position axes spanning ${\sim}1\degr\times1\degr$, and the frequency axis in the double-sideband data 
cubes has 128 channels corresponding to 4400~\kms\ in velocity space. 
The single-band data cube obtained from the new correlator has the same configuration 
except for having 256 frequency channels spanning 9100~\kms. 
These data cubes were made with $w$-projection applied. 
The Briggs robustness parameter was set to zero, and a pixel size of 0.85$\arcsec$ pixel$^{-1}$ was used. 
The synthesised beam size of the final data cubes is ${\sim}3.5\arcsec \times2.4\arcsec$, 
corresponding to ${\sim}16\times11$~kpc$^{2}$ at $z \sim0.32$. 
This high spatial resolution helps to reduce the risk of confusion from companion galaxies, 
which complicates low-resolution {\HI} stacking experiments using a single dish \citep[e.g.,][]{Delhaize:2013}. 
The {\it rms} noise per channel in the double-sideband data cubes is $\sim$153~$\mu$Jy~beam$^{-1}$, 
while the {\it rms} noise level of the single-continuous-band data cube is 190~$\mu$Jy~beam$^{-1}$.

\section{Stacking {\HI} 21-\lowercase{cm} Emission} 
\label{sec:stacking}

\begin{figure*}
 \centering
 \includegraphics[width=150mm]{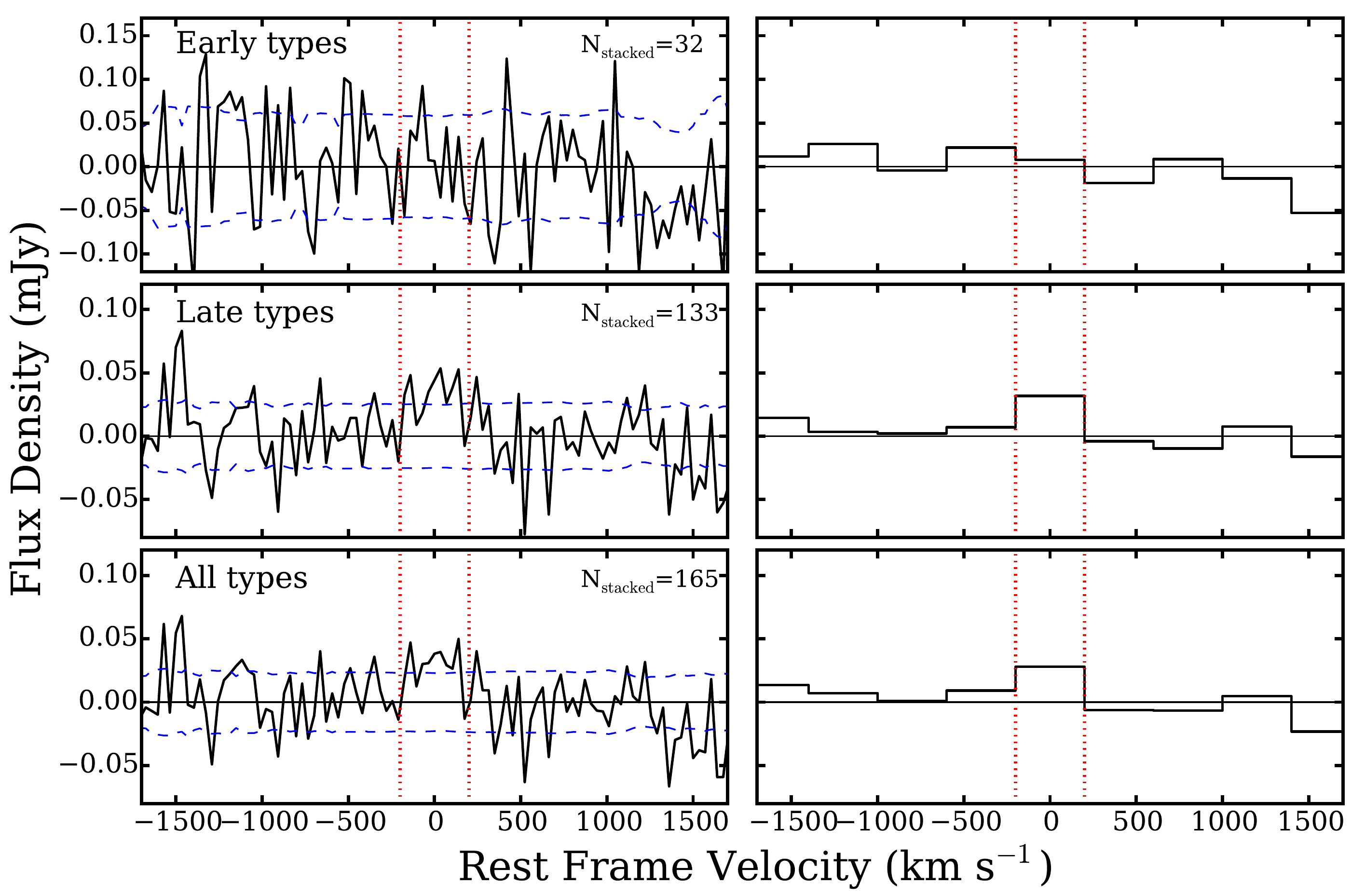}
 \caption{Co-added {\HI} spectra of galaxies in the VVDS~14h field. 
   The {\it left} panels show stacked {\HI} spectra for different galaxy types as classified in Section~\ref{sec:galaxy_class}. 
   The blue horizontal dashed lines show 1$\sigma$ uncertainties of the stacked spectra. 
   The {\it right} panels show the co-added spectra rebinned to 400~{\kms}. 
   The vertical dotted lines in all panels indicate the velocity window in which {\HI} flux was measured and converted to {\HI} mass.} 
 \label{fig:coadded_spectra}
\end{figure*}

\begin{table}
 \caption{{\HI} mass measurements and {\MHI}$/L_{B}$ ratios of the VVDS 14h field galaxies for subsamples.} 
 \centering
 \label{tab:HImass}
 \begin{tabular}{@{}lcccc}
   \hline
         Sample  & $N_{\rm sample}$ & $\langle \MHI \rangle$ & $\langle \MHI \rangle/\langle L_B \rangle$ \\
   \hline 
                      &                     & ($\times$10$^{9} \msun$) & (\msun$/$\lsun)\\
     Early &\032 & 1.56~$\pm$~3.65 (0.4 $\sigma$) & 0.06~$\pm$~0.14\\
     Late  & 133 & 6.49~$\pm$~1.61 (4.0 $\sigma$) & 0.48~$\pm$~0.12\\
     All    & 165 & 5.73~$\pm$~1.47 (3.9 $\sigma$) & -\\
   \hline
 \end{tabular}
\end{table}

\begin{figure}
 \centering
 \includegraphics[width=86mm]{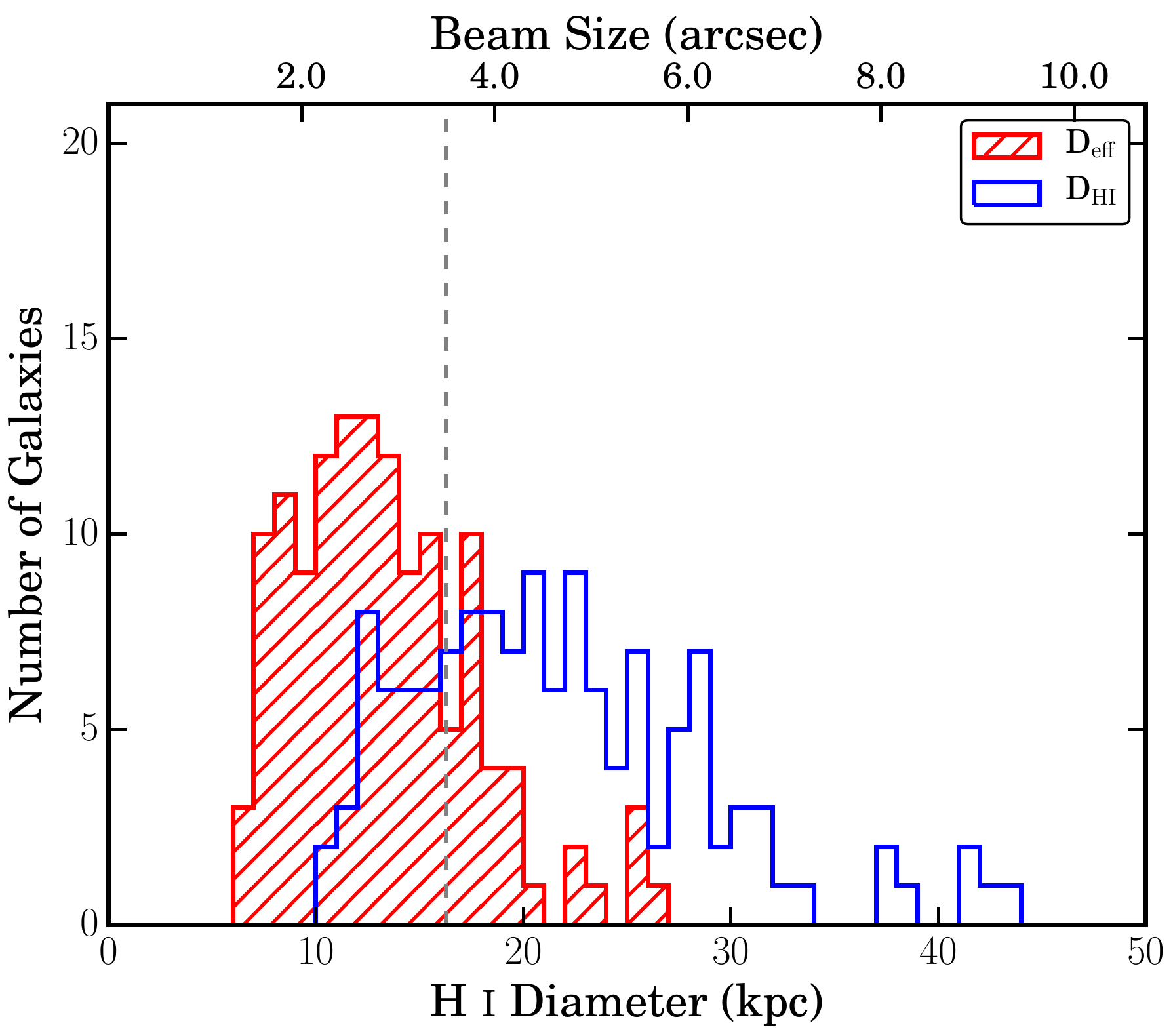}
 \caption{Estimated {\HI} sizes of galaxies in the VVDS 14h field. 
   The {\HI} sizes were derived from the relations between {\HI} size and 
   absolute $B$-band magnitude found by \citet{Broeils_Rhee:1997}. 
   The blue histogram shows the estimated {\HI} diameters at a surface density of 1~\msun~pc$^{-2}$. 
   The red hatched (/) histogram is the distribution of the effective {\HI} diameters of the VVDS~14h galaxies. 
   The vertical dashed line shows the resolution of the GMRT observations.} 
 \label{fig:HIsize}
\end{figure}

\begin{figure}
 \centering
 \includegraphics[width=86mm]{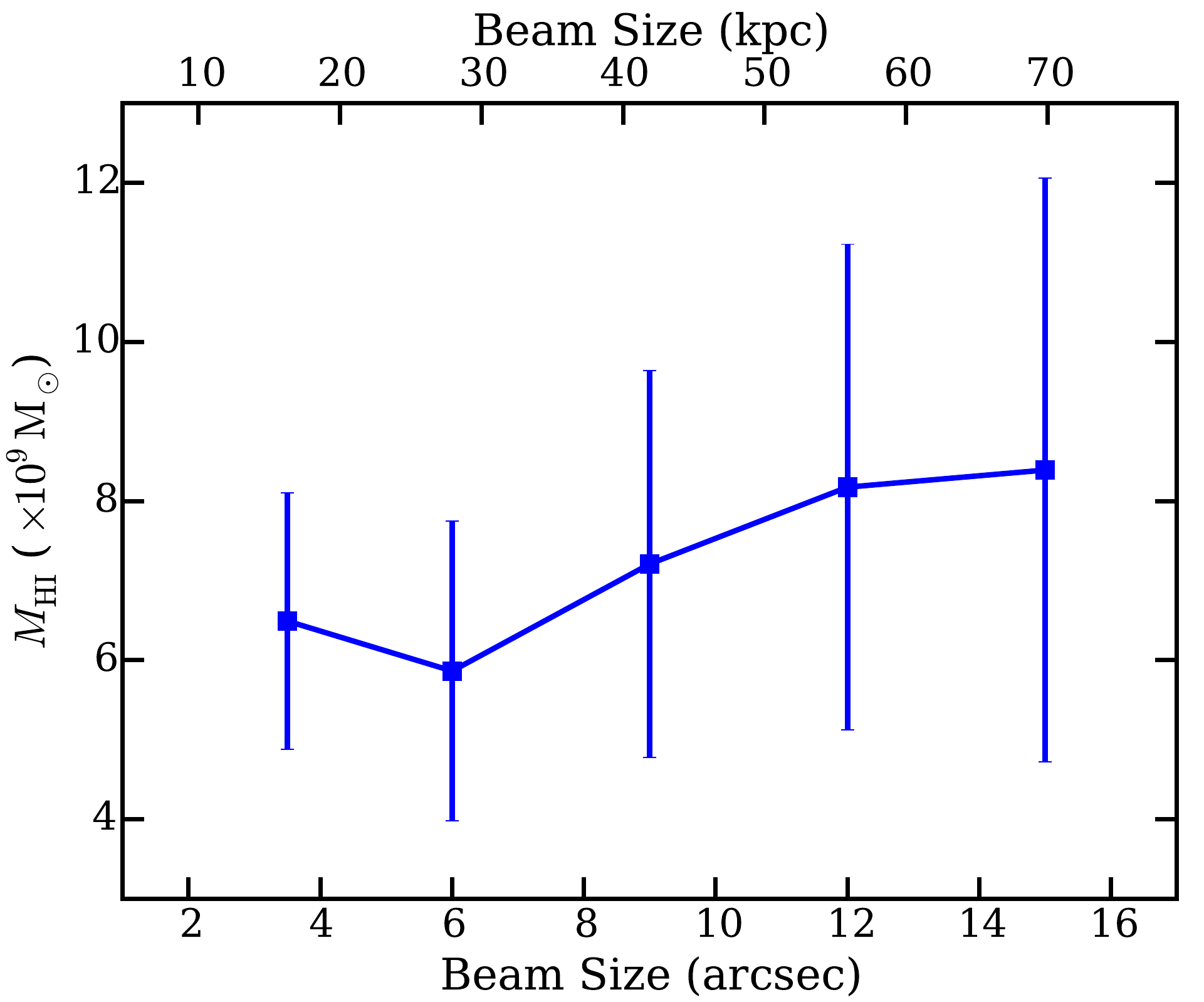}
 \caption{Average {\HI} mass for the VVDS 14h galaxy late-type subsample with different restoring beam sizes.} 
 \label{fig:beam_size_test}
\end{figure}

To make a direct detection of {\HI} emission from an individual galaxy at 5$\sigma$ significance in this GMRT observation, 
the galaxy would have to have an {\HI} mass greater than $\sim$5.7~$\times$~10$^{10}$ {\msun} 
within a 400~{\kms} velocity profile and be smaller than the 3.5$\arcsec$ synthesised beam. 
Such galaxies are rare in the local {\HI} mass function \citep[e.g.,][]{Zwaan:2005,Martin:2010}, 
and it is unlikely that one would fall in the volume surveyed by the GMRT\null. 
Visual inspection of the data cubes yielded no obvious detection of any {\HI} emitters. 
A search using {\sc duchamp} \citep{Whiting:2012}, an automated source-finding routine, 
was carried out through the entire spectral data cubes. 
No direct detection was found in this automated blind search, consistent with the visual inspection.  

To reach an {\HI} emission signal level with increased signal-to-noise ratio, 
spectra from multiple galaxies were co-added, a process called {`\HI} spectral stacking' 
\citep{Chengalur:2001,Zwaan:2001}. 
Redshifts obtained from the MMT/AAT optical spectroscopic observations 
were converted to the corresponding {\HI} frequencies in the data cubes. 
As illustrated in Fig.~\ref{fig:rfi}, some channels of data cubes were severely contaminated by RFI, 
and it was necessary to exclude objects located in these bad channels.
After removing these galaxies from the original sample of 199, 165 
galaxies covering the range $0.3047 \le z\le 0.3370$ remained available for {\HI} stacking.
They are broken into subsamples of 32 early-type and 133 late-type galaxies, respectively. 

The {\HI} spectral stacking followed the methodology described by \citet{Rhee:2013,Rhee:2016}. 
In brief, individual spectra of sample galaxies were extracted and corrected for the primary beam attenuation 
based on the empirically measured
GMRT primary beam pattern.
Then each spectrum was shifted to the galaxy's rest frame velocity before co-adding all the spectra. 
The stacked spectrum was computed from a weighted average using the {\it rms} noise of 
each gain-corrected spectrum to form a weight. 
A stacked spectrum was computed separately for the early-and late-type subsamples 
as well as a grand total spectrum for the full sample of 165 galaxies as plotted in Fig.~\ref{fig:coadded_spectra}.

The co-added spectra for each galaxy type can be converted to {\HI} mass using the following equation \citep{Wieringa:1992}:
\begin{equation}
 \label{eq:HImass}
 \frac{{\MHI}}{\msun} = \frac{236}{(1+z)} \left( \frac{D_L}{{\rm Mpc}} \right)^2 
 \bigg( \frac{\int S_V dV}{{\rm mJy \,km~s^{-1}}} \bigg),
\end{equation} 
where $z$ is redshift, $D_{L}$ is the luminosity distance in units of Mpc, 
and $\int S_V dV$ is the integrated {\HI} emission flux in units of $\rm {mJy~\kms}$. 
To calculate the integrated {\HI} flux in Eq.~\ref{eq:HImass}, 
we assumed the velocity window width of 400~\kms\ within which all {\HI} emission flux was integrated. 
This velocity window was estimated from the Tully-Fisher relation \citep{Tully:1977} 
using $B$ magnitudes converted from the SDSS magnitudes of our sample, 
giving mean $w_{20} =275.5$~\kms\ and maximum 492.1~\kms, where $w_{20}$ is the {\HI} linewidth at 20 per~cent of the peak flux. 
Allowing for redshift uncertainty of the sample galaxies, 
$2\sigma \sim60$~\kms\ along with the velocity profile widths ($w_{20}$) derived from the Tully-Fisher relation, 
the velocity width of 400~\kms\ was used to calculate {\HI} mass from the stacked {\HI} spectra. 
Using different velocity widths from 350~\kms\ to 500~\kms\ makes no systematic difference in the average {\HI} mass.
The mean redshift 0.323 of stacked galaxies was used for this calculation and also for the luminosity distance in Eq.~\ref{eq:HImass}. To estimate the uncertainties of the stacked {\HI} spectra, jackknife resampling \citep{Efron:1982} was adopted.

The average {\HI} mass was calculated separately for early-type and late-type galaxies (see Table~\ref{tab:HImass}). 
The early-type sample is unlikely to have a significant amount of {\HI} gas, 
although the small number of galaxies stacked gives rise to a large statistical uncertainty. 
Most {\HI} gas in this sample seems to reside in the late-type galaxies. 
This  {\HI} distribution can also be seen when comparing the {\HI} gas richness of both populations using the ratio of {\MHI}$/L_{B}$. 
To do this, the $B$-band magnitude was transformed from the SDSS photometry, in particular $g-r$ colour 
and $g$-band magnitude according to the conversion relation provided on the SDSS 
website,\footnote{http://www.sdss.org/dr7/algorithms/sdssUBVRITransform.html} 
and then converted to the $B$-band luminosity after Galactic dust extinction correction and $k$-correction. 
The {\MHI}$/L_{B}$ values are given in Table~\ref{tab:HImass}. 
The {\MHI}$/L_{B}$ ratio shows that the early-type subsample consists of gas-poor galaxies (${\MHI}/{L_B} \la 0.1$) 
as expected for an early-type population \citep{Roberts:1994}. 
In contrast, the late-type subsample shows the typical {\HI} richness (${\MHI}/{L_B} \approx 0.48$) of late-type galaxies. 

In comparison samples at $z \sim0.1$ and 0.2, \citet{Rhee:2013} found 
$\langle \MHI \rangle/\langle L_B \rangle=0.08\pm0.06$ for early-type galaxies 
and $\langle \MHI \rangle/\langle L_B \rangle=0.31\pm0.06$ for late-type galaxies.
The early-type galaxies at $z \sim0.32$ likewise have {\HI} mass-to-light ratios consistent with zero 
although all the measurements have  large uncertainties. 
Late-type galaxies appear to be more gas-rich at $z\sim0.32$ than at $z\sim0.1$--$0.2$ 
although the measurements  overlap within the $1\sigma$ error bars. 
While one may expect an evolutionary trend towards large gas-mass-to-light ratios with increasing redshift, 
the error bars in our measurement are too large to make a definitive claim. 
Further, in comparison to the results from previous large blind {\HI} surveys and an {\HI} stacking analysis with large samples, 
the {\MHI}$/L_{B}$ ratio of our late-type sample does not appear significantly larger. 
Using the HIPASS catalogue with its optical counterparts identified from the 6dF Galaxy Survey \citep[6dFGS,][]{Jones:2004,Jones:2009}, 
\citet{Doyle:2005} investigated the ${\MHI}/L_{B_J}$\footnote{6dFGS measured photometric magnitudes directly from calibrated
images of the SuperCOSMOS Sky Survey \citep{Hambly:2001}. The ${B_J}$ magnitude is a photometric quantity measured from the UK Schmidt IIIa-J photographic plate images of the SuperCOSMOS survey. It is close to the  ${B}$ band in the Johnson--Cousins ${\it UBVRI}$ system: ${B_J} = {B} - 0.230 - 0.237({B-V})$ \citep{Peacock:2016}.}
ratios of a large number of galaxies in the local universe, 
showing the mass-to-light ratio for most of them is 1--2~${\msun}/{\lsun}$. 
This is larger than for our late-type galaxies. \citet{Delhaize:2013} also tested the {\MHI}$/L_{B_J}$ ratio 
of $\sim$18 000 sources using an {\HI} stacking technique. 
Their results have ${\MHI}/L_{\it B_J}$ ratios,
which are in good agreement with \citet{Doyle:2005} 
and decrease with redshifts out to $z \sim$~0.1. 
However, their {\MHI}$/L_{\it B_J}$ at their highest redshift ($z \sim$~0.1) is very similar to the {\MHI}$/L_{B}$ ratio 
that we measure for late types. 
The {\MHI}$/L_{\it B_J}$ measurement of their HIPASS data in an optical luminosity range similar 
to our mean luminosity is consistent with our {\MHI}$/L_{B}$ ratio. 
Therefore the difference of {\MHI}$/L_{B}$ between the large surveys 
and our experiment seems attributable to sample size and selection effects.

To check whether the VVDS 14h galaxies are resolved by the GMRT synthesised beam, 
{\HI} sizes of the sample galaxies were estimated using the relation between optical and {\HI} sizes found by \citet{Broeils_Rhee:1997}. 
They used optical $B$-band photometry and structural properties of nearby galaxies to derive the relations 
between the optical properties and {\HI} size and mass. 
\citeauthor{Broeils_Rhee:1997} defined two kinds of {\HI} diameters related to the absolute $B$-band magnitude. 
One is $D_{\rm HI}$, the {\HI} diameter at a surface density of 1~\msun~pc$^{-2}$, 
and the other is $D_{\rm eff}$, the {\HI} effective diameter enclosing 50$\%$ of the {\HI} mass. 
As seen in Fig.~\ref{fig:HIsize}, most of the late-type sample galaxies (77~per~cent) should be unresolved by the GMRT synthesised beam of $\sim$3.5\arcsec\ in terms of $D_{\rm eff}$, but 75~per~cent of the sample  have $D_{\rm HI}$ larger than the GMRT beam size.
The magnitude--size relations were derived for galaxies within 100~Mpc, but galaxies at $z\approx 0.32$ are generally smaller \citep{vanderwel:2014}.

To check whether the beam size affects the {\HI} measurement, 
the {\HI} mass was re-measured with increasing beam sizes. 
To do this, the original data cubes were smoothed to lower resolutions, 
corresponding to circular Gaussian beams of 6$\arcsec$, 9$\arcsec$, 12$\arcsec$ and 15$\arcsec$, 
i.e., to 28, 42, 56 and 70~kpc, respectively. 
The stacking procedure was then repeated for each of these smoothed cubes, 
and the results are shown in Fig.~\ref{fig:beam_size_test}. 
There is a weak trend for the {\HI} mass measurements to increase with increasing beam size in the late-type galaxies sample.
Part of the increase  may be attributed to increasing confusion with companion gas-rich dwarf galaxies.
This  {\HI} mass increase has low significance, indicating that the {\HI} size of sample galaxies derived in Fig.~\ref{fig:HIsize} 
may be overestimated, perhaps due to the small sample that \citet{Broeils_Rhee:1997} used to derive the relation 
between optical magnitude and {\HI} size.
All in all, there is no indication that gas masses are outside the stated uncertainties of the measurements.

\section{Galaxy Stellar and {\HI} Mass} 
\label{sec:stellar_mass}

\begin{figure}
 \centering
 \includegraphics[width=86mm]{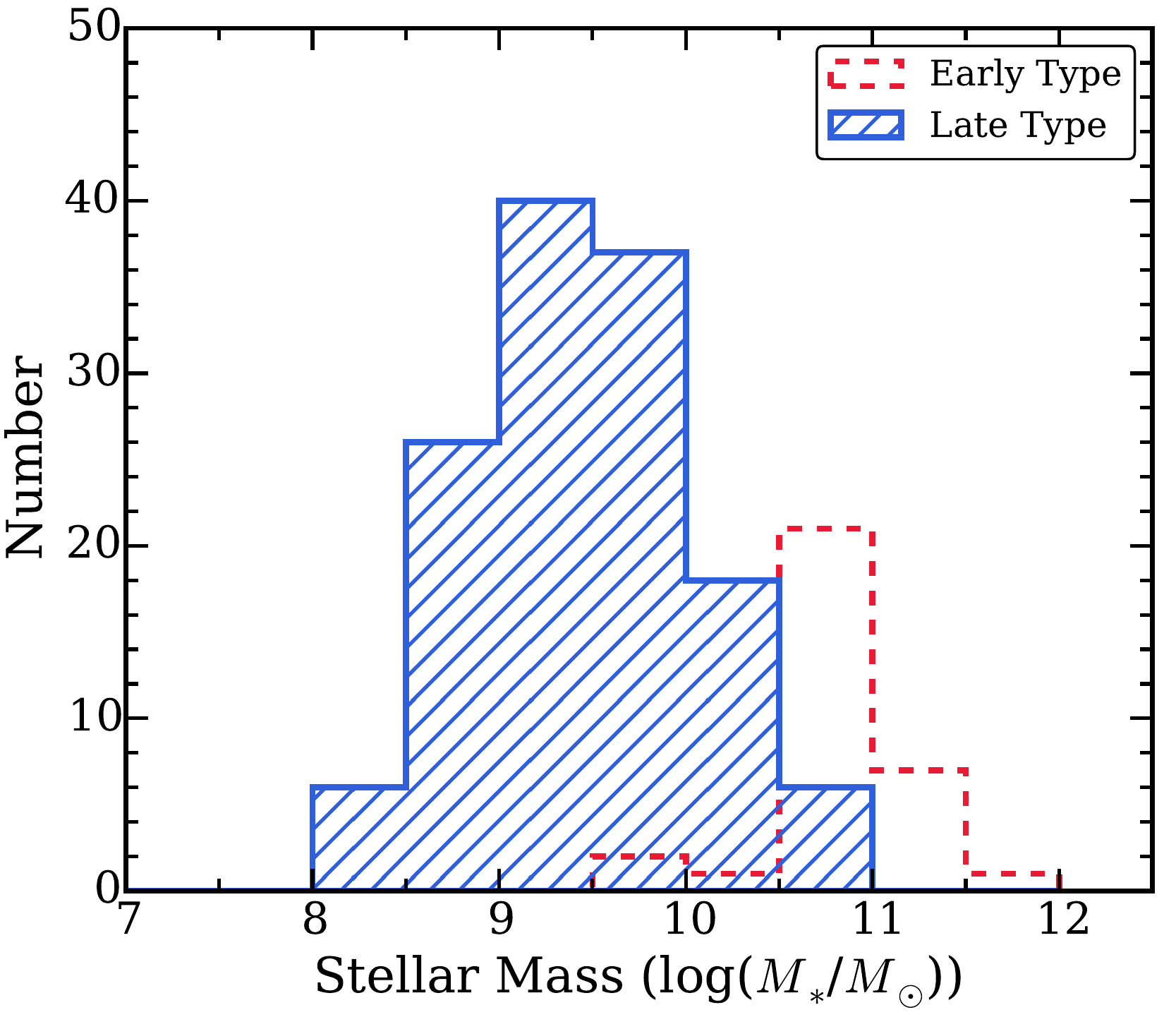}
 \caption{Stellar mass distribution of galaxies used for {\HI} stacking. 
   The blue hatched (/) histogram indicates stellar masses of the late-type galaxies, 
   and early-type galaxies are denoted by the open histogram
   with dashed red outline.}
 \label{fig:smass}
\end{figure}

We determined the stellar mass of each galaxy in the VVDS~14h field sample using the publicly released 
$\chi^{2}$ SED fitting code {\sc le phare}\footnote{{\url {http://www.cfht.hawaii.edu/~arnouts/lephare.html}}} 
\citep{Arnouts:1999,Ilbert:2006}. 
This software was developed to determine galaxy photometric redshifts and  physical parameters, 
for example stellar mass and star formation rate (SFR), by comparing photometry with stellar population synthesis models. 
The code uses the stellar synthesis templates of \citet{Bruzual:2003} with a \citet{Chabrier:2003} initial mass function (IMF) 
to synthesize photometric magnitudes of various bands. 
The 27 models span three metallicities and seven exponentially decreasing star formation profiles 
with $\tau$ = 0.1, 0.3, 1, 2, 3, 5, 10, 15 and 30~Gyr. 
The {\sc le phare} code also applies the extinction law of \citet{Calzetti:2000} 
allowing $E(B-V)$ to range from 0 to 0.6 and stellar population ages from 0 to 13~Gyr. 
Fig.~\ref{fig:smass} shows the stellar mass distribution of our stacked galaxies based on SDSS $ugriz$ magnitudes and our spectroscopic redshifts.
Galaxies with the largest stellar masses are mostly early-types. 
This is as  expected and also shows that the spectroscopic classification is consistent 
with the stellar masses and colours derived from the photometry.

\begin{figure*}
 \centering
 \includegraphics[width=150mm]{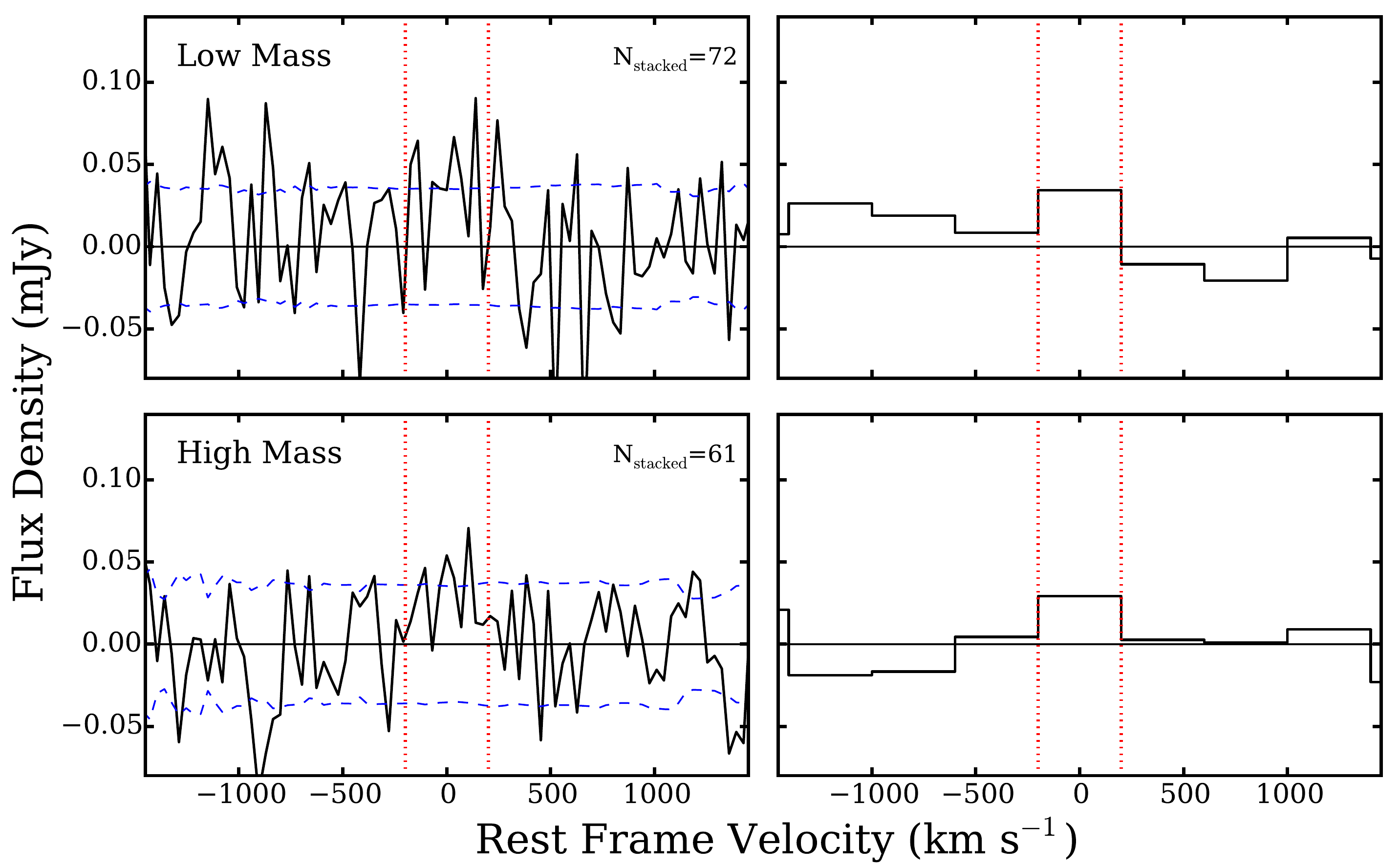}
 \caption{Co-added {\HI} spectra of two subsamples of late-type galaxies defined by stellar masses. 
   The low-mass sample has stellar mass of $M_{*} \leq 10^{9.5}$~{\msun} 
   while the high mass sample contains $M_{*} > 10^{9.5}$~{\msun}.}
 \label{fig:smass_test}
\end{figure*}

\begin{table*}
  \caption{Average {\HI} mass measurements for different stellar mass bins and the fraction of {\HI} and stellar masses.} 
 \centering
 \label{tab:smass}
 \begin{tabular}{@{}lcccc}
   \hline
         Sample  & $N_{\rm sample}$ & $\langle M_{\rm stellar} \rangle$ & $\langle \MHI \rangle$      & $\langle \MHI \rangle / \langle M_{\rm stellar} \rangle$ \\
   \hline 
                     &                          & ($\times$10$^{9} \msun$)       & ($\times$10$^{9} \msun$)  & \\
     Late $M_{*} \leq 10^{9.5}$ \msun   & 72 & 1.24$\pm$ 0.75 & 7.00~$\pm$~2.18 & 5.65 $\pm$ 3.84 \\
     Late $M_{*} >  10^{9.5}$ \msun & 61 & 13.33$\pm$ 15.59 & 5.99~$\pm$~2.45 &  0.45 $\pm$ 0.56 \\
   \hline
 \end{tabular}
\end{table*}

To explore the \HI\ contributions from galaxies with different stellar masses, 
the late-type sample was divided into two subsamples based on the derived stellar mass: 
$M_{*} \leq 10^{9.5}$~{\msun} and $M_{*} > 10^{9.5}$~{\msun}. 
The average {\HI} mass of each subsample was measured by the same {\HI} spectral stacking (see Fig.~\ref{fig:smass_test}). 
Table~\ref{tab:smass} shows that the populations have  similar {\HI} gas amounts, 
but the low-mass sample is dominated by the gaseous component ({\HI}), 
while stellar mass is dominant in the high-mass sample. 
A recent galaxy evolution model \citep{Lagos:2014} predicted this relation between stellar and gaseous mass 
because the galaxy population with $M_{*} \leq 10^{9.5}$~{\msun} has 
its gas mostly in the form of atomic hydrogen (\HI) whereas
the $M_{*} > 10^{9.5}$~{\msun} population has gas mostly in molecular form (H$_{2}$). 
The implication of this relation between stellar and {\HI} mass is 
that the faintness of low-stellar-mass galaxies will bias optically selected {\HI} surveys of neutral gas abundance.

\section{Global {\HI} Gas Density Evolution}

\begin{table*}
 \caption{{\HI} mass density and cosmic {\HI} density. 
   $N_{\rm sample}$ is the number of galaxies that are co-added, 
   $\langle \MHI \rangle$ is the average {\HI} mass (in units of 10$^{9}$~{\msun}), 
   $\langle L_B \rangle$ is the mean $B$-band luminosity in units of 10$^{9}$~{\lsun},
   $\rho_{L_{B}}$ is the luminosity density in units of 10$^{8}$~{\lsun}~Mpc$^{-3}$, 
   {\rhoHI} is the {\HI} density multiplied by the correction factor in units of 10$^{7}$~{\msun}~Mpc$^{-3}$, 
   and {\OHI} is the cosmic {\HI} density.} 
 \centering
 \label{tab:HI_density}
 \begin{tabular}{@{}cccccc}
   \hline
      $N_{\rm sample}$  &  $\langle \MHI \rangle$  & $\langle L_B \rangle$  &  $\rho_{L_{B}}$  &  $\rhoHI$  &  {\OHI}  \\
   \hline
                               &  (10$^{9}$~\msun)  &  (10$^{9}$~\lsun)  &  (10$^{8}$~\lsun~Mpc$^{-3}$)  &  (10$^{7}$~\msun~Mpc$^{-3}$)  &  (10$^{-3}$) \\
      165                  &  5.73~$\pm$~1.47  &  16.20~$\pm$~ 0.11  &  1.92~$\pm$~0.49  &  6.79~$\pm$~2.48  &  0.50~$\pm$~0.18 \\
   \hline
 \end{tabular}
\end{table*}

\subsection{{\HI} Mass Density ({\rhoHI})}

The {\HI} gas density ({\rhoHI}) in the VVDS~14h field
can in principle be calculated from the total {\HI} mass found by the GMRT
divided by the comoving volume surveyed.  However, calculating the effective volume considering survey incompleteness (and to a lesser extent excluded
areas near bright foreground objects) is not trivial.  
The maximum volume available is bounded by the GMRT 10 per~cent beam diameter 
($\sim$0.95~deg) and the GMRT frequency range. 
After accounting for RFI and decreased effective bandpass (Fig.~\ref{fig:rfi}),
this volume (calculated as prescribed by \citealt{Hogg:1999}) is $\sim$41000 comoving Mpc$^{3}$. 
The detected {\HI} density in this volume is 
$(2.31\pm0.60)\times10^{7}$~{\msun}~Mpc$^{-3}$ with
the 133 late-type galaxies giving nearly all of it,
$(2.11\pm0.53)\times10^{7}$~{\msun}~Mpc$^{-3}$. 
The {\HI} contribution of the 32 early-type galaxies is 
$(0.12\pm0.31)\times10^{7}$~{\msun}~Mpc$^{-3}$,
$\la$5 per~cent of the total {\HI} density. While these values are strict lower limits on the {\HI} density in the VVDS~14h field, they are not strict lower limits on the {\em cosmic} {\HI} density because of cosmic variance.   
The original VVDS survey has cosmic variance $<$5\% for $z<0.5$
\citep{Driver:2010}.  However, the narrow velocity range of our survey
reduces the volume and increases the cosmic variance to 
$\sim$54\%.\footnote{\url{http://cosmocalc.icrar.org/}} 

Measuring the actual completeness of our spectroscopic survey is difficult,
especially given the finding that low-mass systems appear to be very hydrogen rich as 
found in Section~\ref{sec:stellar_mass}. 
Our selection criteria were deliberately biased to bright and late-type galaxies in 
order to obtain as many redshifts as possible. 
In order to account for both incompleteness and cosmic variance, we have normalized the 
observed \HI\ density to the blue luminosity ($L_B$) density.  
In effect, we use our measurements of $L(\HI)/L_B$ for the sample galaxies together 
with the known cosmic luminosity density in blue light.
The method, described by \citet{Rhee:2013,Rhee:2016},
relies upon good measurements of the VVDS~14h field in the SDSS $ugriz$ bands. 
While good $ugriz$ photometric measurements are available for the entire galaxy sample, 
the luminosity function and density of the VVDS~14h field have not been computed and made available in the $ugriz$ bands. 
Instead, we used a study by \citet{Ilbert:2005}, who computed the evolution of the galaxy luminosity function 
using the VVDS data in the $UBVRI$ bands. 
\citeauthor{Ilbert:2005} drew on data sets from two fields in the VVDS catalogue, though our VVDS~14h field was not included. 
We used the \citeauthor{Ilbert:2005} luminosity function and density around $z \approx0.32$, 
assuming that variation of the luminosity function and density among the VVDS fields is 
insignificant at the redshift of interest \citep{Garilli:2008}. 
$B$ magnitudes were obtained from the SDSS photometry as described in Section~\ref{sec:stacking} 
and converted to $B$-band luminosities ($L_{B}$) after applying the Galactic dust extinction correction and k-correction.
We then calculated {\rhoHI} for the entire sample applying the volume normalisation as following Eq.~5 of \citet{Rhee:2016}. 
All the values pertaining to this volume normalisation are listed in Table~\ref{tab:HI_density}. 
The effect of the volume correction is to double the {\HI} density. 
An additional correction to {\rhoHI} is needed to account for
galaxies too faint to have been included in our spectroscopic survey. 
The correction factor {$f$} was estimated following the recipe of \citet{Rhee:2013}, which gives $f = 1.41$. 
The final $\rhoHI = (6.79\pm2.48)\times10^7$~{\msun}~Mpc$^{-3}$, 
almost triple the value computed by the simple volume division (see Table~\ref{tab:HI_density}). 

The above method of evaluating {\rhoHI} intrinsically assumes that there is no strong variation of relative gas richness 
(i.e., {\MHI}$/L_{B}$ ratio) over the full $L_B$ range of the galaxy population. 
However, the sample used for this calculation contains both early and late types,
which show different gas richness, because \citet{Ilbert:2005} 
did not divide their sample into two galaxy types. 
Inclusion of early types in estimating the galaxy luminosity function 
changes the luminosity function shape, in particular 
making the faint end slope shallower \citep{Loveday:2012}. 
Our correction for incomplete sampling of the luminosity function 
is sensitive to the slope at the faint end, 
where late-type galaxies with high ${\MHI}/L_{B}$ are found. 
Applying too shallow  a value for this slope means the correction does not add back in enough low-$L_B$ galaxies,
and this means the final {\rhoHI} may still be a little underestimated.  Measuring separate luminosity functions for early- and late-type galaxies could give a more accurate correction.

As mentioned in many {\HI} surveys and {\HI} spectral stacking experiments
\citep[e.g.,][]{Zwaan:1997,Zwaan:2003,Zwaan:2005,Rhee:2013,Rhee:2016},
{\HI} self-absorption has not been accounted for in the {\rhoHI} calculation 
because it is very difficult to assess to what extent {\HI} self-absorption influences the {\HI} mass and mass density
estimation based on even the blind {\HI} survey data \citep{Zwaan:2003,Zwaan:2005}. 
This is due to a lack of reliable measurements required to examine the effect, such as galaxy inclination measured in the optical bands. 
We assume that an underestimate of {\rhoHI} caused by {\HI} self-absorption \citep{Zwaan:1997} 
does not change the main result of this paper although \citet{Braun:2012} asserts the effect increases {\HI} mass 
and density measurements by approximately 34~per~cent. 
However, more data with high spectral and angular resolution are required to determine whether the assertion by \citet{Braun:2012}, based on a small sample, is representative.

\begin{figure*}
 \centering
 \includegraphics[width=160mm]{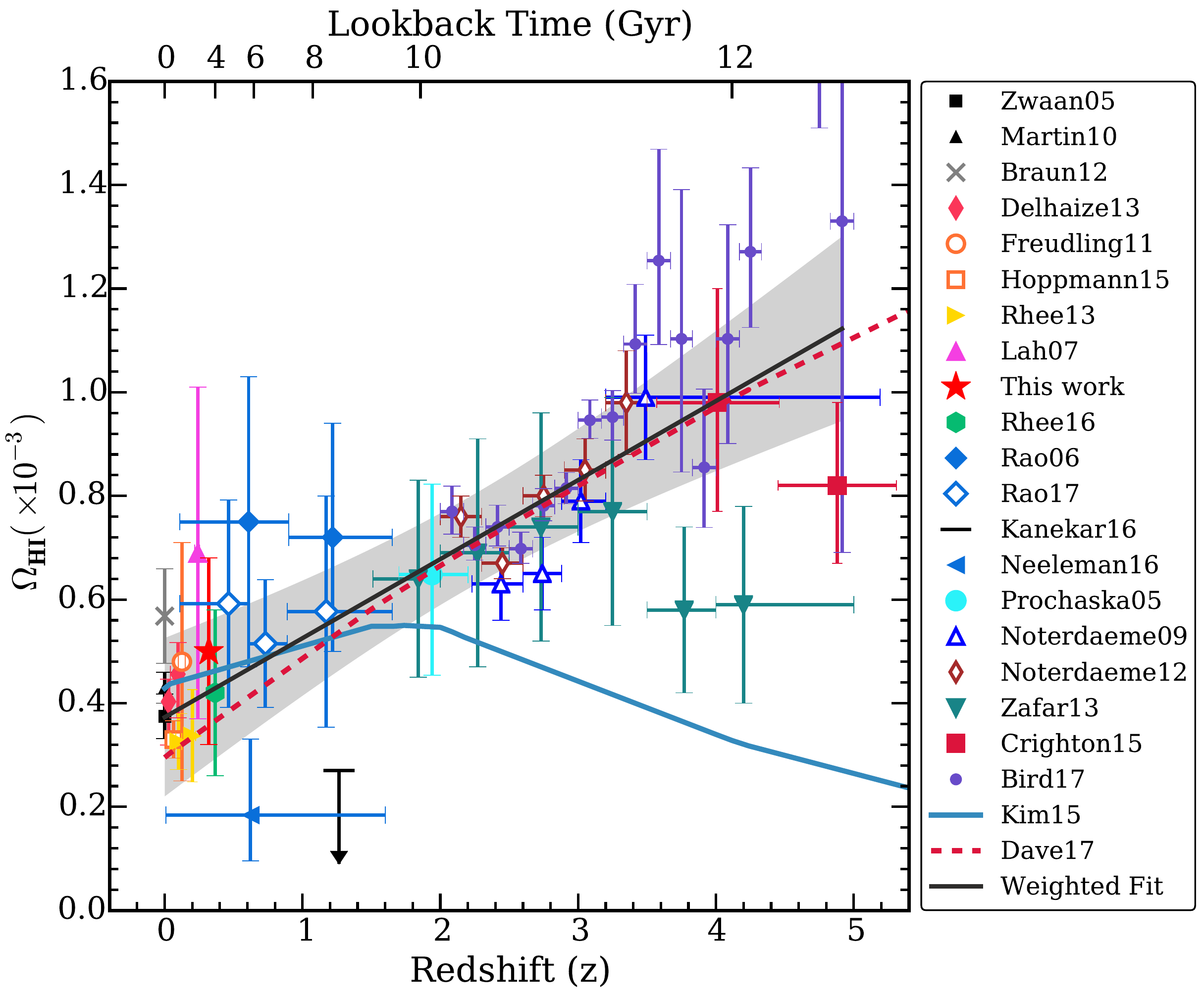}
 \caption{Cosmic {\HI} gas density ({\OHI}) as a function of redshift ({\it bottom} axis) 
   and lookback time ({\it top} axis). All measurements are corrected to the same cosmological parameters. 
   Some DLA measurements adopted a different definition of the cosmic {\HI} density 
   taking into account neutral gas abundance ($\Omega_{\rm gas}$) including helium or contributions 
   from Lyman-$\alpha$ absorbers with lower column density ($\log N({\HI}) < 20.3$). 
   We have corrected all measurements to a consistent definition of {\OHI}. 
   The large red star shows the {\OHI} measurement from this work. 
   The small black square and triangle at $z \sim 0$ are the HIPASS and ALFALFA 21-cm emission
   measurements by \citet{Zwaan:2005} and \citet{Martin:2010}, respectively. 
   The red diamonds are from the Parkes telescope with an {\HI} stacking technique \citep{Delhaize:2013}. 
   The open circle and square are the results from the Arecibo Ultra Deep Survey (AUDS) 
   \citep{Freudling:2011,Hoppmann:2015}. Two right-pointing triangles are measured using the WSRT 
   and {\HI} stacking technique by \citet{Rhee:2013}. The pink triangle is measured by \citet{Lah:2007} 
   applying the GMRT 21-cm emission stacking. The green hexagon denotes the {\HI} stacking measurement 
   for the COSMOS field using the GMRT \mbox{\citep{Rhee:2016}}.
   The black arrow line is the upper limit constrained using the GMRT {\HI} stacking \citep{Kanekar:2016}. 
   The closed and open diamonds, left-pointing triangle, big circle, 
   open triangles, open diamonds and small circles are damped Lyman-$\alpha$ measurements from the HST 
   and the SDSS by \citet{Rao:2006,Rao:2017}, \citet{Neeleman:2016}, \citet{Prochaska:2005}, \citet{Noterdaeme:2009}, 
   \citet{Noterdaeme:2012}, and \citet{Bird:2017}, respectively. 
   The downward triangles and big square at high redshift of $z >$~2 are ESO UVES and Gemini GMOS 
  measurements of DLAs by \citet{Zafar:2013}, \citet{Crighton:2015}, respectively.
  The black line with a shaded area shows a weighted fit of all {\OHI} measurements and its 95 per~cent confidence interval.  
  The solid blue line shows the semi-analytic model prediction of \citet{Kim:2015}, 
  and the dashed red line shows the  {\sc mufasa} \citep{Dave:2017} simulation. } 
 \label{fig:omega_HI}
\end{figure*}

\subsection{The Cosmic {\HI} Mass Density ({\OHI})}

The {\HI} gas density ({\OHI}) in units of the cosmic critical density  is a useful parameter to express 
how the {\HI} gas content of galaxies evolves as a function of cosmic time (redshift). 
This is calculated as the comoving {\HI} gas density scaled by the critical density ($\rho_{\rm crit}$) 
\citep[e.g.,][]{Rhee:2013,Rhee:2016}.
For the VVDS~14h field at $z\approx0.32$, ${\OHI}=(0.50\pm0.18)\times10^{-3}$ 
as seen in Table~\ref{tab:HI_density} and
Fig.~\ref{fig:omega_HI}. 
Other measurements shown in Fig.~\ref{fig:omega_HI}
were obtained using several different techniques: 
{\HI} emission from blind 21-cm emission surveys and {\HI} spectral stacking at the lower redshifts 
\citep{Zwaan:2005,Martin:2010,Freudling:2011,Hoppmann:2015,Delhaize:2013,Rhee:2013,Lah:2007,Rhee:2016};
{\HI} spectral stacking applied to star-forming galaxies at intermediate redshift \citep[$z \approx 1.3$,][]{Kanekar:2016}; 
a targeted survey of {\MgII} absorption systems and a blind DLA survey at intermediate redshifts 
\citep{Rao:2006,Rao:2017,Neeleman:2016}; large DLA surveys at the higher redshifts 
\citep{Prochaska:2005,Noterdaeme:2009,Noterdaeme:2012,Zafar:2013,Crighton:2015,Bird:2017}.
Our {\OHI} measurement agrees within the error bars with all of the measurements made at low redshifts, 
implying that there is little {\HI} gas evolution from $z \sim 0.4$ to present. 
Similarly, at $z >$~2 the {\HI} measurements 
based on large samples of DLAs are also fairly consistent with one another, 
and all indicate an increased {\HI} gas density at $z>2$ 
as seen in Fig.~\ref{fig:omega_HI}.

Despite the agreement at low and high redshifts,
measurements at $0.5< z < 2.0$ are at odds with each other as well as very uncertain. 
This redshift regime is observationally challenging to both 21-cm and DLA techniques. 
At present, measurements in this range come predominantly from DLA studies and have large measurement uncertainties 
and discrepancies. 
The Lyman-$\alpha$ lines used to detect DLAs are still in the UV at these redshifts, 
which means that a space telescope with a UV spectrograph is required.
Furthermore, blind surveys for DLAs are inefficient
because of the low surface density of DLAs at these redshifts. 
Given the limited observing time of {\it HST} for UV spectroscopy, 
no large enough DLA surveys exist. 
In order to increase the efficiency of DLA observations, 
\citet{Rao:2006} adopted pre-selection  based on strong {\MgII} absorption, which most known DLAs have.  
Recently \citet{Neeleman:2016} have compiled  UV spectroscopic data 
from the {\it HST} spectroscopic data archive accumulated over more than 20 years to conduct a blind DLA search. 
Although the number of DLAs that they identified is small (4 DLAs), their result is about four times lower in {\OHI} 
than that of \citet{Rao:2006} with less uncertainty 
and is consistent with the lower-${z}$ {\OHI} measurement from {\HI} 21-cm observations. 
This discrepancy may be interpreted as a bias of the {\MgII} pre-selection method 
(a pre-selection cut of $W_{0}^{\lambda 2796} \geq 0.6$~\AA, the rest-frame equivalent width of the {\MgII} 2796~\AA) 
that \citet{Rao:2006} used \citep{Neeleman:2016,Berg:2017}.  
However, \citet{Rao:2017} recently provided an update of their previous {\MgII} pre-selection DLA survey, increasing their sample size from 41 to 70. They re-derived {\OHI} for a larger sample in the same redshift interval ($0.11 < z < 1.65$). 
As seen in Fig.~\ref{fig:omega_HI}, their {\OHI} measurements are lower than their previous measurements, 
becoming closer to {\HI} measurements at lower redshifts. 
They also claimed that their sample is statistically representative of the true DLA population at $z < 1.65$, denying
claimed bias effects in their pre-selection method. 
Another upper limit on {\HI} gas content was provided by \citet{Kanekar:2016}, who stacked 857 star-forming galaxies at $z=1.265$ in the four DEEP2 fields. Their upper limit is consistent with that of \citeauthor{Neeleman:2016} 
but discrepant with the \citeauthor{Rao:2017} measurements. 
This redshift range has also been explored with a new observing technique called {\HI} intensity mapping 
\citep{Chang:2010,Masui:2013,Switzer:2013}. 
However these measurements are affected by systematics and bias factors that are poorly constrained, resulting in large uncertainty. 
To address the tension in this redshift range, instruments to conduct large blind surveys are likely required.
As mentioned by \citet{Rao:2017}, there seems to be no plan for a large survey for DLA at $z < 1.65$ in the near future.  
However, forthcoming deep {\HI} surveys using the SKA pathfinders will  measure accurate {\OHI} at this redshift even 
before the SKA comes online \citep[e.g., DINGO and LADUMA,][]{Meyer:2009,Holwerda:2012}.
 
The most noticeable trend in Fig.~\ref{fig:omega_HI} is that the DLA measurements at $z > 2$ consistently 
show a larger value of {\OHI} than the {\HI} 21-cm emission measurements at $z < 0.4$. 
Unlike {\HI} gas content measurement using 21-cm emission from galaxies, 
DLAs measure {\HI} gas not only in galaxies but also around galaxies, for example {\HI} streams and filaments. 
The observed difference may be pointing towards an increased contribution of neutral atomic gas outside galaxies at higher redshifts. 
Taking into account all {\OHI} measurements, the overall trend indicates {\OHI} at $z \sim 5$ is at least double the value in the local universe.

Theoretical understanding of the trends seen in Fig.~\ref{fig:omega_HI} remains incomplete, and there are diverse results. 
While cosmological galaxy evolution simulations based on semi-analytic models \citep[e.g.,][]{Lagos:2014,Popping:2014,Kim:2015} are able to match the observed {\HI} gas abundance observed at low redshifts, 
they have had difficulty reproducing the observed abundances at higher redshifts. 
For instance, \citet{Kim:2015} predicted that {\OHI} shows little
evolution at lower redshifts, well reproducing the trend of those {\HI}
observations, but decreases at $z > 2$, in conflict with the trend in
Fig.~\ref{fig:omega_HI}.
One reason for this tension could be that these authors modelled neutral gas only inside galaxies 
without attempting to estimate the neutral gas outside of galaxies, which is hard to model in their simulations.
More recently, a cosmological hydrodynamic simulation called {\sc mufasa}
\citep{Dave:2017} has yielded a good match to the observations (Fig.~\ref{fig:omega_HI}).

\section{Summary and Conclusion}

Stacking 21~cm {\HI} measurements of 165 galaxies at $0.305<z<0.337$ shows 
that $\ga$95~per~cent of the neutral gas is found in blue, star-forming galaxies. 
The average ratio of {\HI} gas mass to blue luminosity in these galaxies (133 in our sample) is ${\sim}0.5$~$\msun/\lsun$. 
Among these late-type galaxies, those having lower stellar mass are more gas-rich than more massive ones. 
The cosmic {\HI} gas density at $z=0.32$ as a fraction of the cosmological critical density is
${\OHI}=(0.50\pm0.18)\times~10^{-3}$. This value is in good agreement with results at $z = 0$ 
based on large-scale 21-cm surveys as well as other {\HI} stacking measurements at $z < 0.4$. 
Our result at $z \approx$~0.32 plays a crucial role in supporting the previous result from \citet{Rhee:2016} 
in the sense that all measurements of {\OHI} at $z < 0.4$ based on 21-cm line emission observations 
are consistent with no evolution in the neutral hydrogen gas density over the last $\sim$4~Gyr. 
However the overall trend including DLA measurements at higher redshifts 
shows a factor of $\ga 2$ higher {\OHI} at $z \sim5$ than at the present epoch.
Observations of {\HI} at $z\ga0.5$ using upcoming radio telescopes 
will better define the evolution of {\OHI} between the local universe, 
where the gas is predominantly inside galaxies, and higher redshifts.

\section*{Acknowledgments}

We are grateful to the anonymous referee for helpful comments on the manuscript.
We thank the staff of the GMRT for their assistance. The GMRT is operated 
by the National Centre for Radio Astrophysics of the Tata Institute of Fundamental Research. 
We are also grateful of the Australian Astronomical Observatory (AAO) as well as 
the staff of MMT Observatory for their assistance. Observations reported here were obtained 
at the MMT observatory, a joint facility of the Smithsonian Institution and the University of Arizona. 
The authors thank Susan Tokarz for help with the MMT/Hectospec data reduction. 
This research was funded by an Australian Indian Strategic Research Fund (AISRF) grant. 
This fund was jointly administered by the Department of Innovation, Industry, Science and Research in Australia 
and by the Department of Science and Technology in India. 
The project title was ``Gas in Galaxies in the Distant Past''. 
Parts of this research were conducted by the Australian Research Council Centre of Excellence for
All-sky Astrophysics (CAASTRO), through project number CE110001020. 
This research uses data from the VIMOS VLT Deep Survey, obtained from the VVDS database operated by Cesam,
Laboratoire d'Astrophysique de Marseille, France. 
Based on data obtained with the European Southern Observatory Very Large Telescope, Paranal, Chile, 
under programs 070.A-9007 and 177.A-0837. 
Funding for the SDSS and SDSS-II has been provided by the Alfred P. Sloan Foundation, the Participating Institutions, 
the National Science Foundation, the U.S. Department of Energy, 
the National Aeronautics and Space Administration, the Japanese Monbukagakusho, 
the Max Planck Society, and the Higher Education Funding Council for England. 
The SDSS Web Site is http://www.sdss.org/. This research made use of {\sc astropy}, 
a community-developed core Python package for Astronomy \citep{Astropy-Collaboration:2013}.

\bibliographystyle{mnras}
\bibliography{VVDS14_reference}

\label{lastpage}
\end{document}